\documentclass[12pt]{article}

\usepackage{amsmath}
\usepackage{amsfonts}
\usepackage{amssymb}
\usepackage{latexsym}
\usepackage{color}
\usepackage[dvips]{graphicx}
\usepackage{cite}
\input{colordvi.tex}

\renewcommand{\thefootnote}{\#\arabic{footnote}}

\begin{document}
\setcounter{footnote}{0}

\begin{titlepage}
\begin{flushright}
RESCEU-43/12

\noindent
ICRR-Report-648-2012-37
\end{flushright}
\begin{center}


\vskip .5in

{\Large \bf
Statistics of general functions of a Gaussian field
-application to non-Gaussianity from preheating-
}
\vskip .45in

{\large
Teruaki Suyama$^{1}$
and 
Shuichiro Yokoyama$^{2}$
}

\vskip .45in%

{\em
$^1$
  Research Center for the Early Universe (RESCEU), Graduate School
  of Science,\\ The University of Tokyo, Tokyo 113-0033, Japan
  }\\
{\em
$^2$
  Institute for Cosmic Ray Research (ICRR),\\
  The University of Tokyo, Kashiwa, Chiba, 277-8582, Japan
  }

\end{center}

\vskip .4in

\begin{abstract}
We provide a general formula for calculating correlators of arbitrary function of a Gaussian field.
This work extends the standard leading-order approximation based on the $\delta N$ formalism 
to the case where truncation of the $\delta N$ at some low order does not yield the correct answer. 
As an application of this formula, we investigate $2$, $3$ and $4$-point functions
of the primordial curvature perturbation generated in the massless preheating model 
by approximating the mapping between the curvature perturbation and the Gaussian field 
as a sum of the many spiky normal distribution functions as suggested by lattice calculations.
We also discuss observational consequences of this case and show that trispectrum would be a
key observable to search signature of preheating in the CMB map.
It is found the forms of the curvature correlation functions for any $\delta N$,
at the leading order in the correlator of the Gaussian field,
coincide with the standard local type ones.
Within this approximation, it is also found that the standard formula for the non-linearity parameters given
by the product of the derivatives of the e-folding number still holds after
we replace the bare e-folding number appearing in the original $\delta N$ expansion 
with the one smoothed in the field space with a Gaussian window function.

\end{abstract}
\end{titlepage}

\renewcommand{\thepage}{\arabic{page}}
\setcounter{page}{1}
\renewcommand{\thefootnote}{\#\arabic{footnote}}

\section{Introduction}
It has become a standard paradigm that the universe experienced 
inflation, the accelerated expansion, in the very early universe.
During inflation, any massless scalar field acquires classical fluctuations
on super-horizon scales originating from sub-horizon scale quantum fluctuations.
One consequence of the inflationary scenario is that one of, or mixture of those
scalar field perturbations convert into the primordial curvature perturbations 
which are then observed as the temperature anisotropy of the Cosmic microwave
background (CMB) and seeds of the large scale structure (for a general review of inflation
and the primordial perturbation, for instance, see \cite{Liddle-Lyth}).
The simplest model to achieve the conversion is to assume that inflaton fluctuations
are solely responsible for the observed curvature perturbations.
Although such a scenario is the simplest, economical and consistent with 
the existing observational data, 
it is possible that more contrived scenarios as exemplified by the curvaton models \cite{Linde:1996gt,Enqvist:2001zp,Lyth:2001nq,Moroi:2001ct} 
and the modulated reheating ones \cite{Dvali:2003em,Kofman:2003nx} are actually realized in our universe.
Therefore, any observable that can help us distinguish the conversion mechanisms
deserves intensive investigation.

In this paper, we will focus on the correlation functions of the curvature perturbation $\zeta$,
especially three-point and higher order correlation functions that describe
the degree of non-Gaussianity of $\zeta$.
Higher order correlation functions have become useful to constrain the 
conversion models because of the fact that the simplest model of inflation yields only 
a negligible amount of non-Gaussianity while large non-Gaussianity up to observable level 
is the generic feature for other models (for example, see \cite{Komatsu:2009kd}).
In particular, combined use of bispectrum and trispectrum can be powerful to constrain
many models that yield large non-Gaussianity \cite{Suyama:2010uj}.
According to the $\delta N$ formalism \cite{Starobinsky:1986fxa,Salopek:1990jq,Sasaki:1995aw,Sasaki:1998ug,Lyth:2004gb}, 
the curvature perturbation on super-horizon
scales is equal to the perturbation of the local e-folding number.
If the expansion of the universe is driven by the scalar fields,  
the perturbation of the e-folding number is produced by the perturbations
of the scalar fields whose evolution can be calculated as if they were
evolving on the homogeneous Friedmann-Lema\^{\i}tre background.
Assuming only a single scalar field sourcing $\zeta$ just for simplicity,
the curvature perturbation at point ${\vec x}$ can be written as a function
of the sourcing scalar field $\chi$ at the same point \cite{Lyth:2005fi};
\begin{equation}
\zeta ({\vec x})=N_1 \chi ({\vec x})+\frac{1}{2!}N_2 \chi^2 ({\vec x})+\frac{1}{3!} N_3 \chi^3 ({\vec x}) +\cdots. \label{intro-series}
\end{equation}
For some models such as the curvaton models, it is a good approximation to treat
the scalar field as the Gaussian variable.
For such a case, the non-Gaussianity of $\zeta$ arises due to the non-linear relation
between $\chi$ and $\zeta$.
In most models that fall into this category, the series expansion above converges so 
rapidly that truncation at the second order is accurate enough to evaluate the three-point
function of $\zeta$.
With this approximation, we can express $\zeta$ in the form of the so-called local type as \cite{Komatsu:2001rj}
\begin{equation}
\zeta=\zeta_g+\frac{3}{5}f_{\rm NL} \zeta_g^2, \label{intro-fnl}
\end{equation}
where $\zeta_g=N_1 \chi$ is the Gaussian part of $\zeta$ and $f_{\rm NL}=\frac{5}{6} \frac{N_2}{N_1^2}$ 
is the non-linearity parameter defining the strength of the non-Gaussianity of $\zeta$ \cite{Lyth:2005fi}.
Current observational limit on $f_{\rm NL}$ by WMAP 9 year data is $-3 < f_{\rm NL} <77$ ($95 \%$ level)
\cite{Bennett:2012fp}.

Although the use of Eq.~(\ref{intro-fnl}) has been quite powerful to constrain
many models that yield non-Gaussian curvature perturbation,
there are several models for which the convergence of the series (\ref{intro-series}) 
is so slow that it does not admit the truncation of it at the first order for the calculation
of the two-point function, at the second order for the calculation of the three-point function, etc.
For instance, in a model where the curvature perturbation is given by trigonometric function like $\zeta \sim \cos (\chi / \Lambda)$,
the expansion (\ref{intro-series}) becomes practically useless when $\langle \chi^2 \rangle \gg \Lambda^2$
(see Sec.~\ref{trigonometric-mapping}).
Another example, which is the main subject of this paper, is the massless preheating
model \cite{Prokopec:1996rr,Greene:1997fu} in which the mapping between $\chi$ and $\zeta$ can be obtained only by means
of the massive numerical calculations.
The chaotic nature of the motions of inflaton and $\chi$ field which is coupled to inflaton
during preheating and termination of the growth of field fluctuations by the highly non-linear
effects defeat accurate analytic derivation of the mapping.
Indeed, the obtained mapping $\zeta (\chi)$ by numerical calculations shows that it is 
quite sensitive to $\chi$, and $\zeta$ suddenly becomes large for some specific values of $\chi$,
which appears as many spikes within the range of $\chi$ given by $\sim \sqrt{\langle \chi^2 \rangle}$. 
Obviously, the expansion of the mapping given by (\ref{intro-series}) around some value of $\chi$
and truncating the series at the lowest order does not work at all for obtaining the correct 
(or accurate) correlation functions of $\zeta$.
The latest evaluation of the mapping $\zeta (\chi)$ was done in \cite{Bond:2009xx}.
In \cite{Bond:2009xx}, the motions of the scalar fields during preheating are solved by 
performing lattice simulation, which automatically includes all the non-linear interactions among 
the scalar fields.
The simulations are done for 11563 different initial values (i.~e.~values just before preheating)
of $\chi$ and the total number of e-fold is read for each simulation to obtain the
mapping $\zeta (\chi)$.
Despite the resultant shape of $\zeta (\chi)$ is far beyond simple polynomials,
the analytic fitting formula expressed by the sum of the normal distribution functions, 
each of which represents the spike, is also provided.
Although they have discussed the possibility of generating the CMB cold spot,
they have not investigated the higher order correlation functions of the resultant primordial
curvature perturbations.
It is then an interesting project to see what kind of forms the correlation functions of $\zeta$
take (especially, two-point and three-point functions) as a result of the (almost) chaotic 
mapping in preheating, which motivates our paper.
In Ref. \cite{Kohri:2009ac}, 
the authors have also investigated the amplitude of the primordial non-Gaussianity from preheating
by using a smoothed quadratic function for $\zeta (\chi)$.
Our paper differs from \cite{Kohri:2009ac} in that we obtain the exact correlation
functions of $\zeta$ by adopting the analytic fitting formula presented in \cite{Bond:2009xx} 
without applying smoothing procedure from the outset.
We also show that use of the smoothed e-folding number in the field space
as the expansion coefficients of the $\delta N$ expansion (\ref{intro-series}) is
correct at the leading order in the correlation function of $\chi$.

In this paper, we first provide a basic formulation of calculating
the correlation functions of $\zeta (\chi)$ \cite{David-Middleton}.
By making use of this formula, we can either numerically or analytically
evaluate the primordial non-Gaussianity in a feasible manner,
for cases in which the truncation of Eq.~(\ref{intro-series}) at the lowest order is 
not necessarily justified.
An important assumption behind our formulation is the Gaussianity of the sourcing
scalar field $\chi$. 
After we explicitly verify that our formulation correctly recovers the standard local form 
of the three-point function for $\zeta$ given by Eq.~(\ref{intro-fnl}),
we then apply our formulation to the models mentioned above such as the one where
the curvature perturbation is given by a trigonometric function
and the massless preheating scenario.
For this purpose, we will adopt the fitting formula given by \cite{Bond:2009xx} as the 
correct mapping for the massless preheating.
For the physically relevant scales, it will turn out that the three-point and the
four-point functions from preheating still take the standard local forms. 
The difference from the standard one appears in how the non-linearity parameters are
related to the quantities characterizing $\zeta (\chi)$.
It is found that $f_{\rm NL}$ from preheating is typically enormous ${\cal O}(10^6)$.
We also find that the curvature perturbation from preheating itself does not have enough amplitude
to explain the observed amplitude, which by necessity, 
requires that the total curvature perturbation is a mixture of dominant component 
and preheating component which is subdominant.
Taking the dominant component to be the standard adiabatic Gaussian perturbation from inflaton,
this mixture dilutes the non-Gaussianity of the total curvature perturbation and,
as a result, $f_{\rm NL}$ becomes typically ${\cal O}(0.1)$, 
which is below the observational sensitivity.
More noticeable signal appears in $\tau_{\rm NL}$, one of the non-linearity parameters
characterizing the strength of a part of the four-point function \cite{Byrnes:2006vq}, 
which is boosted typically up to ${\cal O}(10^2)$.
These findings suggest that observational study of the trispectrum is 
a key to dig up the trace of preheating left in the curvature perturbation. 

In section \ref{gf}, we provide a general formalism to evaluate the correlation function
of $\zeta$ which is not necessarily written as the perturbative expansion form like Eq.~(\ref{intro-fnl}).
After demonstrating the effectiveness of the formalism by applying it to some simple
models in section \ref{ex-1},
detailed analysis for the case of preheating is developed in section \ref{ex-2}.
In section \ref{Gc}, we show that, by means of the diagrammatic approach, 
reduction to the standard local type non-Gaussianity
seen for the preheating case is a generic feature that holds for other models.
The last section is conclusion.

\section{General formalism}
\label{gf}
Let us provide a general formalism to evaluate the correlation function of
the curvature perturbation which is not necessarily written as 
the perturbative expansion form \cite{David-Middleton}.
Our primary assumption is that the curvature perturbation $\zeta$ in real space
is a function of a Gaussian scalar field $\chi$ at the same point:
\begin{equation}
\zeta (\vec{x})=f ( \chi ({\vec x}))-\langle f ( \chi ({\vec x})) \rangle. \label{def-zeta}
\end{equation}
The second term is introduced so that $\langle \zeta \rangle$ is zero.
Without a loss of generality, we can set $\langle \chi \rangle=0$
(if $\langle \chi \rangle \neq 0$, we can redefine $\chi$ by subtracting
$\langle \chi \rangle$ from the original field.).
For our purpose, it is convenient to introduce the Fourier transformation
of $f(\chi)$ as \footnote{The authors appreciate Jun'ichi Yokoyama for suggesting this transformation.}
\begin{equation}
f (\chi)=\int \frac{d\sigma}{2\pi} ~f_\sigma e^{i \chi \sigma},~~~~~
\Longrightarrow ~~~~~
f_\sigma=\int d\chi~f(\chi) e^{-i \chi \sigma}.
\end{equation}
Using $f_\sigma$ given above, $\langle f (\chi) \rangle$ becomes
\begin{equation}
\langle f (\chi) \rangle =\int \frac{d\sigma}{2\pi} ~f_\sigma 
\langle \exp \left( i \chi ({\vec x}) \sigma \right) \rangle.
\end{equation}
For the Gaussian field $\chi$, we have the following relation,
\begin{equation}
\bigg\langle \exp \left( \int d^3x ~b({\vec x}) \chi ({\vec x}) \right) \bigg\rangle
=\exp \bigg[ \frac{1}{2} \int d^3 x_1 d^3 x_2~\int \frac{d^3 q}{{(2\pi)}^3} P_\chi (q) 
e^{i {\vec q} \cdot ({\vec x_1}-{\vec x_2})} b ({\vec x_1}) b ({\vec x_2}) \bigg],
\end{equation}
for arbitrary function $b({\vec x})$.
Here, $P_\chi(k)$ is the power spectrum of the scalar field $\chi$, which is defined by
\begin{equation}
\langle \chi_{\vec k}\chi_{\vec k'} \rangle ={(2\pi)}^3 P_\chi (k) \delta ({\vec k}+{\vec k'}),
\end{equation}
where $\chi_{\vec k}$ is defined by 
$\chi_{\vec k}=\int e^{-i{\vec k}\cdot {\vec x}}\chi ({\vec x})d^3x$. 
Using this formula, we find
\begin{equation}
\langle f (\chi) \rangle =\int \frac{d\sigma}{2\pi} ~f_\sigma e^{ -\frac{\langle \chi^2 \rangle}{2}\sigma^2 }, \label{<f>}
\end{equation}
where $\langle \chi^2 \rangle \equiv \langle \chi^2 ({\vec x}) \rangle$.
Once $f_\sigma$ is given, we can calculate
$\langle f (\chi) \rangle$ by performing one-dimensional integral.

In a similar way, we can write down the formal expression of the $N$-point function
of $f$ as
\begin{eqnarray}
\langle f ({\vec x_1}) \cdots f({\vec x_N}) \rangle
&=&\int \frac{d\sigma_1 \cdots d\sigma_N}{{(2\pi)}^N}~f_{\sigma_1} \cdots f_{\sigma_N} 
\langle e^{i\chi ({\vec x_1}) \sigma_1+ \cdots +i \chi ({\vec x_N}) \sigma_N} \rangle \nonumber \\
&=&\int \left( \prod_{i=1}^N \frac{d\sigma_i}{2\pi}~f_{\sigma_i} 
e^{-\frac{\langle \chi^2 \rangle}{2} \sigma_i^2} \right)
\exp \left( -\langle \chi^2 \rangle \sum_{i<j} \sigma_i \sigma_j \xi_\chi (r_{ij}) \right), \label{N-point}
\end{eqnarray}
where $\xi_\chi$ is defined by 
$\xi_\chi (r_{ij})=\langle \chi ({\vec x_1}) \chi ({\vec x_2}) \rangle /\langle \chi^2 \rangle$ and
$r_{ij}=|{\vec x_i}-{\vec x_j}|$.
Again, once we know $f_\sigma$, we can in principle calculate the $N$-point function
by performing $N$-dimensional integral.

Eq.~(\ref{N-point}) is applicable to any power spectrum of $\chi$ field
as long as $\chi$ field is Gaussian.
For the scale invariant power spectrum which we will assume hereafter,
$P_\chi$ is given by
\begin{equation}
P_\chi (k)=\frac{P_0}{k^3},~~~~~~~~~P_0:{\rm constant}.
\end{equation}
Then, we have
\begin{equation}
\langle \chi^2 \rangle =\int \frac{d^3 k}{{(2\pi)}^3} P_\chi (k) 
=\frac{P_0}{2\pi^2} \int_{L^{-1}}^{q_{\rm max}} \frac{dk}{k}=\frac{P_0}{2\pi^2} \ln (q_{\rm max}L).
\end{equation}
Here, both UV and IR cutoffs are introduced.
The IR cutoff corresponds to the box size $\sim L$ in which perturbation is defined
and UV cutoff is dependent on the model we consider.
Carrying out the similar calculation for $\langle \chi ({\vec x}) \chi ({\vec y}) \rangle$,
we finally arrive at
\begin{equation}
\xi_\chi (r)=\frac{-C_i (r/L)+C_i (q_{\rm max}r)+\frac{\sin (r/L)}{r/L}
-\frac{\sin (q_{\rm max}r)}{q_{\rm max}r}}{\ln (q_{\rm max}L)}, \label{scale-inv-xi}
\end{equation}
where $C_i (x)$ is the cosine integral function defined by
\begin{equation}
C_i (x) = -\int_x^\infty \frac{\cos t}{t} dt.
\end{equation}

\section{Application to some simple models}
\label{ex-1}

Here, we present the application of our formula provided in the previous section
to some simple models.

\subsection{Case of quadratic mapping}
For many models including curvaton model and modulated reheating model
that produce non-Gaussian curvature perturbation,
it is a good approximation to truncate the polynomial of $\zeta (\chi)$ at
the second order for the evaluation of the leading contribution to the three-
point function:
\begin{equation}
f(\chi)=a \chi+b \chi^2,~~~~~a,b:~{\rm constants.} \label{quadra}
\end{equation}
Although it is much simpler to calculate the correlation functions directly from
the above equation as is widely done in the literature, 
let us here apply the formalism derived in the previous section just in order
to verify explicitly that the formalism indeed yields the same results as the standard one.
The Fourier transform of the quadratic expression above is written as
\begin{equation}
f_\sigma=2\pi \left( ia \frac{d}{d\sigma}-b \frac{d^2}{d\sigma^2} \right) \delta (\sigma).
\end{equation}
Plugging this into Eq.~(\ref{<f>}) and performing the integration by parts, 
we find
\begin{equation}
\langle f({\vec x}) \rangle =\langle \chi^2 \rangle \int d\sigma ~\delta (\sigma)
( -ia  \sigma+b ) e^{-\frac{\langle \chi^2 \rangle}{2} \sigma^2}=b \langle \chi^2 \rangle,
\end{equation}
which is obvious from direct operation of $\langle \cdots \rangle$ onto both sides
of Eq.~(\ref{quadra}).
In a similar way, the two-point function of $\zeta$ becomes
\begin{eqnarray}
\langle \zeta({\vec x}) \zeta({\vec y}) \rangle &=&
\int d\sigma_1 d\sigma_2 ~\delta (\sigma_1) \delta (\sigma_2) 
\left( -ia \frac{d}{d\sigma_1}-b \frac{d^2}{d\sigma_1^2} \right) \nonumber \\
&&\times \left( -ia \frac{d}{d\sigma_2}-b \frac{d^2}{d\sigma_2^2} \right)
e^{-\frac{\langle \chi^2 \rangle}{2} (\sigma_1^2+\sigma_2^2)-\langle \chi^2 \rangle \sigma_1 \sigma_2 \xi_\chi (r)} -b^2 \langle \chi^2 \rangle^2 \nonumber \\
&=&a^2 \langle \chi^2 \rangle \xi_\chi (r)+2b^2 \langle \chi^2 \rangle^2 \xi_\chi^2 (r),
\end{eqnarray}
which also coincides with the result based on the standard calculation.
For the three-point function, after some cumbersome calculation, we arrive at
\begin{equation}
\langle \zeta ({\vec x_1})\zeta ({\vec x_2})\zeta ({\vec x_3}) \rangle=
2a^2 b \langle \chi^2 \rangle^2 (\xi_\chi (r_{12}) \xi_\chi(r_{13})+2~{\rm perms.})
+8b^3 \langle \chi^2 \rangle^3 \xi_\chi (r_{12}) \xi_\chi (r_{23}) \xi_\chi (r_{13}). \label{quadr-3}
\end{equation}
When $\zeta$ is sourced only by a single field as is considered in this paper,
it is common to write $\zeta$ as a sum of the gaussian part and its square as is given by Eq.~(\ref{intro-fnl}).
Identifying $\zeta_g$ as $\chi$ and setting $a=1,~b=\frac{3}{5}f_{\rm NL}$ in Eq.~(\ref{quadr-3}),
we have
\begin{equation}
\langle \zeta ({\vec x_1})\zeta ({\vec x_2})\zeta ({\vec x_3}) \rangle=
\frac{6}{5}f_{\rm NL} ( \langle \zeta ({\vec x_1}) \zeta ({\vec x_2}) \rangle\langle \zeta ({\vec x_2}) \zeta ({\vec x_3}) \rangle
+2~{\rm perms.}) + {\cal O}(\langle \zeta^2 \rangle^3), \label{3-fnl}
\end{equation}
at the leading order in $\zeta_g$.

\subsection{Case of trigonometric function mapping}
\label{trigonometric-mapping}
As the next example which is more non-trivial than the first one, 
let us consider a case where $f(\chi)$ is a sine function:
\begin{equation}
f(\chi)=A \sin \left( \frac{\chi}{\Lambda} \right), \label{sin-map}
\end{equation}
where $A$ and $\Lambda$ are constants.
If $\langle \chi^2 \rangle \ll \Lambda^2$ is satisfied,
the argument of the sine function becomes much smaller than unity
and the Taylor expansion of the sine function up to the lowest order 
required to obtain the non-vanishing correlation function works well \footnote{Precisely speaking, since $\chi$ is a statistical variable obeying
Gaussian distribution, it is possible that $\chi$ takes a value much
larger than $\Lambda$ at some domain of the universe even when $\langle \chi^2 \rangle \ll \Lambda^2$.
But its probability is highly suppressed and practically it does not matter
to assume $\chi /\Lambda \ll 1$ from the inception, that is, 
before taking the statistical average.}.
For instance, $\sin (x) \approx x$ is sufficient to obtain the approximate 
form of the two-point function of $\zeta$.
On the other hand, for the case of $\langle \chi^2 \rangle \gg \Lambda^2$,
Taylor expansion up to the lowest order does not provide a correct answer and 
it is such a case in which the formalism derived in this paper becomes useful.
For the moment, we do not assume any magnitude relationship between
$\langle \chi^2 \rangle$ and $\Lambda^2$.
The Fourier transform of $f$ is given by the superposition of the $\delta$-functions:
\begin{equation}
f_\sigma=A \int d\chi~\sin \left( \frac{\chi}{\Lambda} \right) e^{-i\chi \sigma}=-\pi i A 
\left( \delta( \sigma-\Lambda^{-1})-\delta (\sigma+\Lambda^{-1}) \right).
\end{equation}
Using this equation, we find that $\langle f({\vec x}) \rangle=0$,
which can be also understood from the fact that $\zeta$ is an odd function of $\chi$.
The two-point function then becomes \footnote{In \cite{Yamamoto:1992zb},
two-point function of the baryon isocurvature perturbations having the same form as Eq.~(\ref{sin-map}) was studied.}
\begin{equation}
\langle \zeta ({\vec x}) \zeta ({\vec y}) \rangle=
A^2 e^{-\frac{\langle \chi^2 \rangle}{\Lambda^2}} \sinh \left( \frac{\langle \chi^2 \rangle}{\Lambda^2} \xi_\chi (r) \right).
\end{equation}
We see that $\xi_\chi \ll 1$ is not sufficient to allow the truncation of the Taylor expansion
at a few lowest order when $\langle \chi^2 \rangle \gg \Lambda^2$.
In fact, a stronger condition,
$\xi_\chi (r) \ll  \Lambda^2/ \langle \chi^2 \rangle \ll 1$, needs to be satisfied.

Again plugging the above $f_\sigma$ to Eq.~(\ref{N-point}) and setting $N=3$,
we can verify that the three-point function vanishes exactly.
This should be so since the product of the three $\zeta$s is an odd function of $\chi$.
It is straightforward to extend this result to the $N$-point function
of $\zeta$ where $N$ is any odd number and to conclude that it exactly vanishes.

In a similar way, let us consider the case of cosine function given by
\begin{eqnarray}
f(\chi) = A \cos \left( {\chi \over \Lambda}\right).
\end{eqnarray}
The Fourier transform of $f$ is given by
\begin{eqnarray}
f_\sigma = \pi A \left( \delta (\sigma-\Lambda^{-1}) + \delta (\sigma + \Lambda^{-1})\right).
\end{eqnarray}
Contrary to the sine function, since a cosine function is even,
$N$-point function of $\zeta$ becomes non-zero even where $N$ is odd number.
Actually, we have
\begin{eqnarray}
\langle f({\vec x})\rangle = A e^{- {\langle \chi^2 \rangle \over 2\Lambda^2}},
\end{eqnarray}
and 
\begin{eqnarray}
\langle \zeta({\vec x}) \zeta ({\vec y} )\rangle
= A^2 e^{- {\langle \chi^2 \rangle \over \Lambda^2}} \left[ 
\cosh \left( {\langle \chi^2 \rangle \over \Lambda^2} \xi_\chi(r)\right) - 1
\right].
\end{eqnarray}
Then, the 3-point function of $\zeta$ is 
\begin{eqnarray}
\langle \zeta({\vec x}_1)\zeta({\vec x}_2)\zeta({\vec x}_3)\rangle
&=& A^3 e^{-{3 \langle \chi^2 \rangle \over 2 \Lambda^2}}
\Biggl\{
\Biggl[ 
\cosh \left( {\langle \chi^2 \rangle \over \Lambda^2} \xi_\chi (r_{12}) \right)
\cosh \left( {\langle \chi^2 \rangle \over \Lambda^2} \xi_\chi (r_{23}) \right)
\cosh \left( {\langle \chi^2 \rangle \over \Lambda^2} \xi_\chi (r_{31}) \right) \cr\cr
&&
-
\sinh \left( {\langle \chi^2 \rangle \over \Lambda^2} \xi_\chi (r_{12}) \right)
\sinh \left( {\langle \chi^2 \rangle \over \Lambda^2} \xi_\chi (r_{23}) \right)
\sinh \left( {\langle \chi^2 \rangle \over \Lambda^2} \xi_\chi (r_{31}) \right) 
\Biggr] \cr\cr
&& 
- \Biggl[
\cosh \left( {\langle \chi^2\rangle \over \Lambda^2} \xi_\chi(r_{12})\right)\cr\cr
&& \qquad\qquad
+
\cosh \left( {\langle \chi^2\rangle \over \Lambda^2} \xi_\chi(r_{23})\right)
+
\cosh \left( {\langle \chi^2\rangle \over \Lambda^2} \xi_\chi(r_{31})\right)
\Biggr] + 2 \Biggr\}.
\end{eqnarray}
Like a sine function mapping, 
the condition $\xi_\chi \ll 1$ is not sufficient to perform the Taylor expansion up to some
lowest order in this cosine mapping case, too.
The above general expression is much complicated, but 
for the case with $\langle \chi^2\rangle \xi_\chi(r)/\Lambda^2 \ll 1$
we find simple relations given as
\begin{eqnarray}
\langle \zeta({\vec x}) \zeta ({\vec y}) \rangle
\propto \xi_\chi(r)^2
\end{eqnarray}
and
\begin{eqnarray}
\langle \zeta({\vec x}_1)\zeta({\vec x}_2)\zeta({\vec x}_3)\rangle
\propto \xi_\chi(r_{12})\xi_\chi(r_{23}) \xi_\chi(r_{31}).
\end{eqnarray}
This limiting case corresponds to that the curvature perturbation is a quadratic function of $\chi$
without a linear term, that is, $\zeta = \chi^2$.

\section{Application to massless preheating}
\label{ex-2}
From the illustrations above, we have seen that the formula (\ref{N-point}) can
actually work to evaluate the correlation function of $\zeta$.
Now, let us focus on the case of preheating (for a general review of preheating,
see \cite{Bassett:2005xm}), which is the main part of this paper.
If the test scalar field $\chi$ couples to the inflaton $\phi$ by the interaction:
\begin{equation}
{\cal L}=-\frac{g^2}{2} \phi^2 \chi^2,
\end{equation}
parametric resonance generically occurs after the inflaton starts the oscillations \cite{Traschen:1990sw,Kofman:1994rk,Kofman:1997yn}.
In particular, for inflaton potential given by the quartic $\frac{\lambda}{4}\phi^4$, 
so-called massless preheating scenario \cite{Prokopec:1996rr,Greene:1997fu},
the fluctuations of the $\chi$ field are not suppressed on super-horizon scales 
at the end of inflation when $g^2/\lambda \sim 2$ \cite{Bassett:1999cg,Finelli:2000ya}.
Interestingly enough, the resonance band completely covers the super-horizon scale modes
for $g^2/\lambda=2$ \cite{Greene:1997fu} and thus the $\chi$ field fluctuations grow exponentially on
super-horizon scales,
which has invoked intensive studies of whether the $\chi$ field perturbation
in the massless preheating model
can generate the curvature perturbation on super-horizon scales during 
preheating \cite{Bassett:1999mt,Bassett:1999cg,Finelli:2000ya,Tsujikawa:2000ik,Zibin:2000uw,Tsujikawa:2002nf,Nambu:2005qh,BasteroGil:2007mm}
(for the similar analysis in the other preheating models, see, for instance, \cite{Enqvist:2005qu,Enqvist:2005nc,Hazra:2012kq}). 
Linear perturbation analysis (linear both in metric and field perturbations) shows that the curvature perturbation $\zeta$ on super-horizon scales
grows exponentially at the first stage of preheating \cite{Bassett:1999cg,Finelli:2000ya} \footnote{
A term ``super-Hubble'' instead of ``super-horizon'' is used in these papers because of the fact
that all the perturbation modes we consider are actually in the causal horizon stretched significantly by inflation.
Nevertheless, following convention widely used in the literature, we use ``super-horizon'' in this paper.}.
Effects of the backreaction from the field perturbations by means of the Hartree approximation
were included in \cite{Tsujikawa:2000ik,Zibin:2000uw}.
It was found that the backreaction terminates the growth of $\zeta$.
Second order perturbation including the metric perturbations was applied to the massless
preheating in \cite{Jokinen:2005by} and it was found that the non-Gaussianity also grows 
during preheating (although their procedure to calculate $f_{\rm NL}$ was shown to be inappropriate
in \cite{Kohri:2009ac}).

In \cite{Tanaka:2003cka}, $\delta N$ formalism \footnote{The word $\delta N$ formalism
was not as popular as nowadays and a term ``separate universe approach'' 
was used in the paper.},
which takes into account the fully non-linear effects of the super-horizon dynamics,
was applied to the massless preheating to evaluate the correspondence between 
$\zeta$ and $\chi$ (the function $f(\chi)$) for the first time.
This was done by solving numerically the background equations of motion of the scalar fields 
on the Friedmann-Lema\^{\i}tre universe.
The spatial fluctuations of the scalar fields, which become important at the non-linear stage
of preheating, are not taken into account in the analysis.
It was observed that $\delta N$ is quite sensitive to the change of initial value of $\chi$.
The origin of the acute sensitivity was identified as the (almost) chaotic behavior
of the equations of the scalar fields.
Thus $f(\chi)$ can hardly be approximated by a quadratic expression, 
contrary to many other cases where the truncation of $f(\chi)$ at 
quadratic or cubic order is enough to evaluate the bispectrum or trispectrum
of the non-Gaussian curvature perturbation. 
The same conclusion was obtained in \cite{Suyama:2006rk} in which calculations of $\delta N$
were performed for more number of different initial values of $\chi$.
It was also suggested super-horizon scale curvature perturbations are generated due to the
imperfect randomness of the mapping $f(\chi)$.

Lattice calculations which automatically include the inhomogeneities of the scalar fields 
as well as the fully nonlinear interactions were implemented in \cite{Chambers:2007se} (see also \cite{Chambers:2008gu}).
From the resulting mapping $f(\chi)$, $f_{\rm NL}$ was estimated by fitting $f(\chi)$
with the smoothed quadratic formula (\ref{intro-fnl}).
The latest lattice calculations done by a different simulation code called DEFROST \cite{Frolov:2008hy}
show that within the field range $\sqrt{\langle \chi^2 \rangle}$,
a lot of spikes having different amplitudes look to appear randomly and
uniformly in logarithmic interval of $\chi$.
Although the actual form of $f(\chi)$ given in \cite{Bond:2009xx} is quite complex and chaotic
at a first glance, the analytic approximation that describes the basic behavior
of $f(\chi)$ is proposed, which is given by the sum of the normal distribution:
\begin{equation}
f (\chi)=\sum_p A_p \exp \left( -\frac{{(\chi-\chi_p)}^2}{2\kappa_p^2} \right), \label{fitting-f}
\end{equation}
where $\chi_p$ and $\kappa_p$ represent the position and
the width of the $p$-th spike, respectively.
There are two remarks worth mentioning at this moment. 
First, since $\chi$ is defined such that $\langle \chi \rangle =0$,
$\chi$ used in this paper is different from the one appearing in the original Lagrangian
or the one used in \cite{Bond:2009xx} by a constant, which we denote by $\chi_0$.
This quantity represents the contribution from the long wavelength modes larger than the box 
size $\sim L$ in which our perturbations are defined and effectively works at a part of the
background value in our box. 
Due to the statistical properties of fluctuations,
$\chi_0$ also varies as a stochastic variable as we shift the position of the box in larger space.
Thus, apart from the variance $\langle \chi_0^2 \rangle$, we cannot predict a definite 
value of $\chi_0$ and should treat $\chi_0$ as a free parameter.
Later, we will study the dependence on $\chi_0$ of the non-Gaussianity 
of the curvature perturbation from preheating.
Secondly, in the figure of $f(\chi)$ given in \cite{Bond:2009xx}, we observe two different
types of spikes, i.~e.~, the spikes that have relatively large amplitudes and large widths
and look to appear ${\cal O}(10)$ times as we change $\chi$ by one order of magnitude,
and the ones that have much lower amplitudes and smaller widths and appear much more
frequently (almost continuously).
All the spikes in the former category have positive peaks, that is, $A_p >0$,
while the ones in the latter case have both positive and negative peaks, which appear
to occur with approximately the equal probability.
As a result, $f(\chi)$ becomes positively large near the vicinity of any spike 
belonging the first category and fluctuates finely around zero outside them. 
In the following calculation, we will make an approximation that only the spikes in the first
category contribute to Eq.~(\ref{fitting-f}).
Since each spike enter the correlators of $\zeta$ in the combination $A_p \kappa_p$,
which will be shown later, 
the spikes in the second category would less contribute compared to the 
spikes in the first category because of the smallness of their amplitudes
and partial offset among them.

\subsection{A mock $f(\chi)$}

Let us first consider to generate a mock $f(\chi)$ that mimics the one obtained in Ref. \cite{Bond:2009xx}.
We take this approach by the two reasons.
The first one is that we do not have the precise raw numerical data of \cite{Bond:2009xx} but only have 
numbers derived from the Fig.~1 of \cite{Bond:2009xx} by using the 
public software \footnote{http://www.frantz.fi/software/g3data.php}
which extracts the coordinates out of the graph.
We will see in \ref{comparison} that results using the mock $f(\chi)$ show a good agreement with
the ones obtained from the Fig.~1 of \cite{Bond:2009xx} using the method mentioned above.
The second reason, which is more fundamental, is that we also want to study
how sensitively the slight change of $f(\chi)$, {\it i.~e.} amplitudes, positions
of spikes and their widths, changes the resultant correlation functions of $\zeta$
and to derive the generic properties of the curvature perturbation having such $f(\chi)$ given by Eq.~(\ref{fitting-f}).
\begin{figure}[t]
  \begin{center}{
    \includegraphics[scale=1.7]{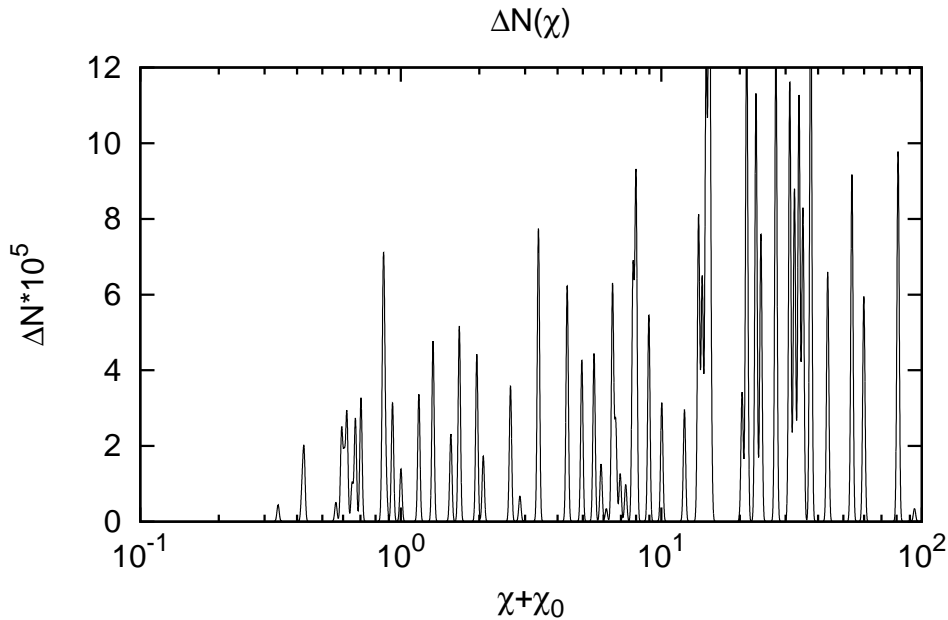}
    }
  \end{center}
  \caption{A mock $f(\chi)$ randomly generated according to the probability
  distribution explained in the text. The parameters are such that 
  $\chi_0=10$ and $\langle \chi^2 \rangle=10$
   in a unit of $10^{-7}M_{\rm Pl}$.}
 \label{deltaN-chi}
\end{figure}
Due to many spikes whose positions apparently appear randomly in $f(\chi)$, 
as was observed in \cite{Bond:2009xx},
we decide to generate the positions of the spikes $\chi_p$ randomly
which are uniformly distributed in $\ln (\chi+\chi_0)$.
Following Ref. \cite{Bond:2009xx}, we set the range of $\chi+\chi_0$  
to be $0.1 = \chi_{\rm min} \le \chi+\chi_0 \le \chi_{\rm max}=100$.
Hereinafter, we present values of $\chi$ in a unit of $10^{-7} M_{\rm Pl}$.
The number of spikes within this interval is fixed to be $75$
and the spikes are randomly generated within this interval.
We have checked that the results are insensitive to both the changes 
of $\chi_{\rm min}$ and $\chi_{\rm max}$ provided that 
$\chi_0+\sqrt{\langle \chi^2 \rangle}$ is well below $\chi_{\rm max}$.
This weak dependence of the results on $\chi_{\rm max}$ is due to the
exponentially suppressed probability of realizing $\chi+\chi_0 > \chi_{\rm max}$.
Although this does not generically hold for the shift of $\chi_{\rm min}$,
the weak dependence on $\chi_{\rm min}$ arises due to the lower and thinner spikes
for smaller $\chi+\chi_0$. 
We also assign the amplitude $A_p$ by an equation
$A_p=\alpha |\log_{10} (10(\chi_p+\chi_0))|^{1.3}$,
where $\alpha \in (0,~4.5\times 10^{-5})$ is chosen randomly for each spike,
and $\kappa_p$ by an equation $0.01 (\chi+\chi_0)$.
These equations are employed to yield a mock $f(\chi)$ that closely resembles the original one,
apart from the fine spikes mentioned earlier.
Because of the probabilistic procedure, we obtain different but similar $f(\chi)$
by each realization.  
By this approach, we are able to see how the resulting 
non-linearity parameters vary by each realization and by the different
choice of the parameters such as $\chi_0$ and $\langle \chi^2 \rangle$. 
As an illustration, in Fig.~\ref{deltaN-chi}, we show a mock $f(\chi)$ for 
a typical case in which $\chi_0=10$ and $\langle \chi^2 \rangle=10$ .
The result actually looks similar to the original $f(\chi)$ provided in \cite{Bond:2009xx}.


\subsection{Two-point function}

Since $f (\chi)$ is given by the sum of the normal distribution,
we can immediately derive the analytic from of $f_\sigma$,
which is given by
\begin{equation}
f_\sigma=\sqrt{2\pi} \sum_p A_p \kappa_p \exp \left( -\frac{\kappa_p^2 \sigma^2}{2}-i\chi_p \sigma \right).
\label{f_sigma-preheating}
\end{equation}
Then, using Eq.~(\ref{<f>}), it can be easily checked that
$\langle f(\chi) \rangle$ is given by
\begin{equation}
\langle f(\chi) \rangle =\sum_p A_p \frac{\epsilon_p}{\sqrt{1+\epsilon_p^2}} 
\exp \left( -\frac{\eta_p^2}{2(1+\epsilon_p^2)} \right), \label{1-point}
\end{equation}
where $\epsilon_p$ and $\eta_p$ are respectively the peak width and the peak position normalized by
$\sqrt{\langle \chi^2 \rangle}$ defined as
\begin{equation}
\epsilon_p \equiv \frac{\kappa_p}{\sqrt{\langle \chi^2 \rangle}},~~~~~
\eta_p \equiv \frac{\chi_p}{\sqrt{\langle \chi^2 \rangle}}.
\end{equation}
Using the definition of $\zeta$ given by Eq.~(\ref{def-zeta}),
the two-point function of $\zeta$ becomes
\begin{equation}
\langle \zeta ({\vec x}) \zeta ({\vec y}) \rangle 
= \langle f ({\vec x}) f ({\vec y}) \rangle -{\langle f(\chi) \rangle}^2. \label{2-point-1}
\end{equation}
The second term on the right hand side is already given by Eq.~(\ref{1-point}).
The first term, by
substituting Eq.~(\ref{f_sigma-preheating}) to Eq.~(\ref{N-point}) for $N=2$, can be written as
\begin{eqnarray}
\langle f ({\vec x}) f ({\vec y}) \rangle &=&
\int \frac{d\sigma_1 d\sigma_2}{4\pi^2} \zeta_{\sigma_1} \zeta_{\sigma_2} 
\exp \bigg[ -\frac{1}{2} \langle \chi^2 \rangle (\sigma_1^2+\sigma_2^2)-\langle \chi ({\vec x}) \chi ({\vec y}) \rangle \sigma_1 \sigma_2 \bigg] \nonumber \\
&=&\sum_{p_1,p_2} \frac{A_{p_1}A_{p_2} \epsilon_{p_1}\epsilon_{p_2}}{\sqrt{(1+\epsilon_{p_1}^2)(1+\epsilon_{p_2}^2)-\xi_\chi^2 (r)}} \nonumber \\
&&\times \exp \left( -\frac{1}{2} \frac{(1+\epsilon_{p_1}^2) \eta_{p_1}^2+(1+\epsilon_{p_2}^2) \eta_{p_2}^2-2\xi_\chi(r) \eta_{p_1} \eta_{p_2}}{(1+\epsilon_{p_1}^2) (1+\epsilon_{p_2}^2)-\xi_\chi^2 (r)} \right).
\label{2-point-2}
\end{eqnarray}
Let us consider a situation where $\langle \chi^2 \rangle \gg \kappa_p^2$.
Eq.~(\ref{2-point-2}) then can be approximately written as
\begin{equation}
\langle f ({\vec x}) f ({\vec y}) \rangle \approx \frac{1}{\sqrt{1-\xi_\chi^2 (r)}} \sum_{p_1,p_2}
A_{p_1}A_{p_2}\epsilon_{p_1}\epsilon_{p_2} \exp \bigg[ -\frac{\eta_{p_1}^2+\eta_{p_2}^2-2\eta_{p_1}\eta_{p_2} \xi_\chi (r)}{2(1-\xi_\chi^2 (r))} \bigg]. \label{2-point-3}
\end{equation}
Since terms in the summation of Eq.~(\ref{2-point-3}) are exponentially suppressed
for $|\eta_p| \gtrsim 1$, 
the summation is mostly contributed by the terms having $|\eta_p| \lesssim 1$.
This means that, contrary to the sine function case considered in the previous subsection,
$\xi_\chi \ll 1$ is the sufficient condition to perform the Taylor expansion of 
Eq.~(\ref{2-point-3}) in terms of $\xi_\chi$ and to truncate
the series at the non-trivial lowest order that contains the non-trivial 
information of the correlator of $\zeta$.
For the case of the two-point function, it turns out that we need to expand
up to first order in $\xi_\chi$,
\begin{equation}
\langle f ({\vec x}) f ({\vec y}) \rangle =
I_0^2+I_1^2 \xi_\chi(r)+{\cal O}(\xi_\chi^2), \label{2-point-4}
\end{equation}
where $I_0$ and $I_1$ are defined by \footnote{
Although not written explicitly, both $I_0$ and $I_1$ depend on $\chi_0$.}
\begin{equation}
I_0=\sum_p A_p \epsilon_p e^{-\frac{\eta_p^2}{2}},~~~~~
I_1=\sum_p A_p \epsilon_p \eta_p e^{-\frac{\eta_p^2}{2}}. \label{def-I0}
\end{equation}
Applying again the approximation $\langle \chi^2 \rangle \gg \kappa_p^2$
to Eq.~(\ref{1-point}), we find 
\begin{equation}
\langle f(\chi) \rangle =I_0.
\end{equation}
It is clear that the first constant term in Eq.~(\ref{2-point-4})
exactly cancels with the second term in Eq.~(\ref{2-point-1}).
In the diagrammatic language, the first term and the second term 
in Eq.~(\ref{2-point-4}) correspond to the disconnected and the connected 
diagram, respectively, and the non-trivial information are contained in
the connected diagram.
Then two-point function of $\zeta$ becomes
\begin{equation}
\langle \zeta ({\vec x}) \zeta ({\vec y}) \rangle 
=I_1^2 \xi_\chi(r)+{\cal O}(\xi_\chi^2). \label{2-point-5}
\end{equation}
This shows that in the regime where $\xi_\chi \ll 1$,
the two-point function of $\zeta$ is proportional to $\xi_\chi$ and the scale dependence of $\xi_\chi$
is stored in $\langle \zeta(\vec{ x}) \zeta(\vec{ y}) \rangle$.

\begin{figure}[t]
  \begin{center}{
    \includegraphics[scale=1.5]{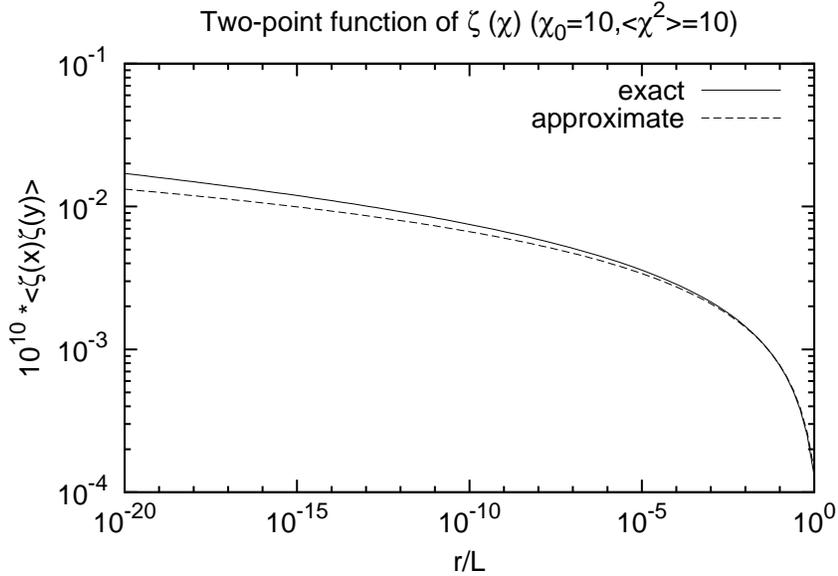}
    }
  \end{center}
  \caption{Two-point function of $\zeta(\chi)$ for one realization 
  ($I_0=0.67\times 10^{-5},~I_1=0.19\times 10^{-5},~I_2=0.46\times 10^{-5}$). 
  The solid line represents the exact
  correlation function calculated from Eqs.~(\ref{1-point}) and (\ref{2-point-2}).
  The dashed line represents the approximated correlation function given by
  Eq.~(\ref{2-point-5}). In both cases, the correlation functions are decreasing
  function. At the left edge where $r/L=10^{-20}$, we find $\xi_\chi \simeq 0.38$ and
  it decreases to $\xi_\chi \simeq 0.017$ at $r/L=1$. }
 \label{two-point-function}
\end{figure}

Fig.~\ref{two-point-function} shows the two-point function of $\zeta(\chi)$
for the same parameter choice, $\chi_0=10,~\langle \chi^2 \rangle =10$.
The solid line is the exact correlation function calculated from 
Eqs.~(\ref{1-point}) and (\ref{2-point-2}).
The dashed line represents the approximated correlation function given by
Eq.~(\ref{2-point-5}) in which second and higher order terms in $\xi_\chi$ are dropped. 
Here, we have assumed $q_{\rm max} L = 10^{53}$  which appears in $\xi_\chi( r )$.
This is because
$q_{\rm max}$ is roughly equal to the Hubble scale at the end of inflation
for the massless preheating case with $g^2 /\lambda=2$ and
taking $L$ to be the size of the present Universe
we have $q_{\rm max} L \sim 10^{53} \left( \frac{H_{\rm end}}{10^{13}{\rm GeV}}\right)$.
At $r/L=10^{-20}$, $\xi_\chi$ becomes as large as $0.38$ and it monotonically
decreases to $0.017$ at $r/L=1$.
The current observations can probe the primordial fluctuation from the current Horizon
scale down to the several order of magnitude smaller than it.
This leads to the huge hierarchy $L \gtrsim r \gg q_{\rm max}^{-1}$.
For such an observationally relevant range, $\xi_\chi (r)$ given by 
Eq.~(\ref{scale-inv-xi}) becomes mildly smaller than unity (typically $0.05 \sim 0.1$).
Although $\xi_\chi$ is suppressed only mildly in this case, 
in order to capture the basic point of our result in analytic way,
we regard $\xi_\chi$ as the expansion parameter and Taylor-expand the correlators with respect to it
as can be seen in Fig.~\ref{two-point-function}.

It is interesting to consider the meaning of the first term on the R.~H.~S. of
Eq.~(\ref{2-point-5}).
Using the definition of $I_1$ given by Eq.~(\ref{def-I0}),
we have
\begin{eqnarray}
I_1=\frac{1}{\langle \chi^2 \rangle} \sum_p A_p \kappa_p \chi_p 
e^{-\frac{\chi_p^2}{2 \langle \chi^2 \rangle}} 
=\frac{1}{\sqrt{2\pi}\langle \chi^2 \rangle} \int d\chi'~f(\chi'+\chi_0) \chi' e^{-\frac{\chi'^2}{2 \langle \chi^2 \rangle}}.
\end{eqnarray}
We now introduce a Gaussian window function $W_g (\chi)$ defined by
\begin{equation}
W_g (\chi)=\frac{1}{\sqrt{2\pi \langle \chi^2 \rangle}}e^{-\frac{\chi^2}{2 \langle \chi^2 \rangle}}. \label{gauss-window}
\end{equation}
This satisfies the normalization condition $\int d\chi~W_g (\chi)=1$
and effectively cuts off contributions from $|\chi| \gtrsim \sqrt{\langle \chi^2 \rangle}$.
With the Gaussian window function, $I_1$ can be written as
\begin{equation}
I_1=\frac{1}{\sqrt{\langle \chi^2 \rangle}}
\int d\chi'~f(\chi'+\chi_0) \chi' W_g(\chi'). \label{I1-gauss}
\end{equation}
On the other hand, we can define $f^R(\chi)$ \footnote{
This quantity was computed numerically in \cite{Suyama:2006rk, Bond:2009xx} 
to estimate the resulting curvature perturbation on super-horizon scales.}
which is defined as
a smoothed $f(\chi)$ in the region 
$(\chi-\sqrt{\langle \chi^2 \rangle},\chi+\sqrt{\langle \chi^2 \rangle})$
by
\begin{equation}
f^R (\chi)=\int d\chi'~f(\chi+\chi') W_g (\chi'). \label{def-fR}
\end{equation}
We can verify that $I_1$ is related to the first derivative of $f^R (\chi)$
evaluated at $\chi=\chi_0$,
namely,
\begin{equation}
I_1 = \sqrt{\langle \chi^2 \rangle} \frac{df_\chi^R}{d\chi}\bigg|_{\chi=\chi_0}.
\end{equation}
Therefore, we can rewrite Eq.~(\ref{2-point-5}) as
\begin{equation}
\langle \zeta ({\vec x}) \zeta ({\vec y}) \rangle 
\approx {\left( \frac{df_\chi^R(\chi_0)}{d\chi} \right)}^2 
\langle \chi({\vec x}) \chi({\vec y}) \rangle.
\end{equation}
This result shows that even when the original e-folding number is spiky and
cannot be approximated as linear expression of the scalar field at all,
at the leading order in $\xi_\chi$
the two-point function of $\zeta$ is given by the first derivative
of the e-folding number, just like the standard $\delta N$ formula,
provided it is smoothed over the range $\sim \sqrt{\langle \chi^2 \rangle}$.
As we will see later, this correspondence holds for the three-point function
and even for the $N$-point function of a general function $f(\chi)$
(at leading order in $\xi_\chi$).

\subsection{Non-Gaussianity}
It is a straightforward and standard calculation to perform the 
three-dimensional integral of Eq.~(\ref{N-point}) for $f_\sigma$
given by Eq.~(\ref{f_sigma-preheating}).
However, since the resulting expression is complex and long,
we write down the expression after applying the approximation 
$\langle \chi^2 \rangle \gg \kappa_p^2$ as what we did for the case
of the two-point function.
The result is given by
\begin{equation}
\langle f ({\vec x_1}) f ({\vec x_2}) f ({\vec x_3}) \rangle 
=\sum_{p_1,p_2,p_3} \frac{A_{p_1}\epsilon_{p_1}A_{p_2}\epsilon_{p_2}A_{p_3}\epsilon_{p_3}}{\sqrt{S}}
\exp \left( -\frac{T}{2S} \right),
\end{equation}
where $S$ and $T$ are defined by
\begin{eqnarray}
S&=&1-\xi_\chi^2 (r_{12})-\xi_\chi^2 (r_{23})-\xi_\chi^2 (r_{31})
+2 \xi_\chi (r_{12}) \xi_\chi (r_{23}) \xi_\chi (r_{31}), \\
T&=&\eta_1^2+\eta_2^2+\eta_3^2-2 \{ \eta_1 \eta_2 (\xi_\chi (r_{12})-\xi_\chi (r_{31}) \xi_\chi (r_{23}))
+2~{\rm perms.} \} \nonumber \\
&&-\eta_1^2 \xi_\chi^2 (r_{23})-\eta_2^2 \xi_\chi^2 (r_{23})-\eta_3^2 \xi_\chi^2 (r_{12}),
\end{eqnarray}
where $r_{ij} = |{\vec x}_i - {\vec x}_j|$.
In the regime where $\xi_\chi \ll 1$, the three-point function of $\zeta$ is
\begin{eqnarray}
\langle \zeta({\vec x}_1) \zeta({\vec x}_2) \zeta({\vec x}_3)\rangle
&=&(I_2-I_0)I_1^2 \left( \xi_\chi(r_{12}) \xi_\chi(r_{13}) 
+ 2~{\rm perms.} \right) + {\cal O}(\xi_\chi^3). \nonumber \\
&\approx& \frac{I_2-I_0}{I_1^2} \left( \langle \zeta ({\vec x_1}) 
\zeta ({\vec x_2}) \rangle \langle \zeta ({\vec x_1}) 
\zeta ({\vec x_3}) \rangle+2~{\rm perms.} \right), \label{3-point-2}
\end{eqnarray}
where $I_2$ is defined by
\begin{equation}
I_2 = \sum_p A_p \epsilon_p \eta_p^2 e^{-\frac{\eta_p^2}{2}}.
\end{equation}
Interestingly, the last equation takes the same form as Eq.~(\ref{3-fnl}),
i.~e.~, the three-point function of $\zeta$ at physically relevant scales
becomes the standard local type one common to many models such as curvaton model
and the modulated reheating one.
The corresponding $f_{\rm NL}$ is given by 
\begin{equation}
\frac{6}{5}f_{\rm NL}=\frac{I_2-I_0}{I_1^2}. \label{ex-fnl}
\end{equation}
In the same way as we derived Eq.~(\ref{I1-gauss}), we can write
$I_0$ and $I_2$ in terms of the Gaussian window function as
\begin{equation}
I_0=\int d\chi'~f(\chi'+\chi_0) W_g (\chi'),~~~
I_2 =\frac{1}{\langle \chi^2 \rangle} \int d\chi'~f(\chi'+\chi_0) \chi'^2 W_g (\chi').
\end{equation}
We can show that the second derivative of $f^R (\chi)$ 
defined by Eq.~(\ref{def-fR}) is given by
\begin{equation}
\frac{d^2 f^R (\chi_0)}{d\chi^2} =\frac{1}{\langle \chi^2 \rangle}
(I_2-I_0).
\end{equation}
Then, we have 
\begin{equation}
\frac{6}{5}f_{\rm NL}=\frac{f^R_{\chi \chi}}{{(f^R_\chi)}^2}. \label{ex-fnl-fR}
\end{equation}
Therefore, $f_{\rm NL}$ is also given by the standard expression
based on the $\delta N$ formalism provided the e-folding number is understood
as the smoothed one over the region $\sim \sqrt{\langle \chi^2 \rangle}$\footnote{
By adopting values of $f^R_\chi$ and $f^R_{\chi \chi}$ provided by \cite{Bond:2009xx},
this fact was used in \cite{Kohri:2009ac} to compute $f_{\rm NL}$ from the massless
preheating.}.

Given the current situation where cosmic measurements turn out to be
useful to probe the four-point function, or trispectrum (for instance, see \cite{Komatsu:2010hc}),
it is intriguing to go one step further from the three point function 
and to evaluate the four-point function.
The calculation is tedious but straightforward.
Again with the approximations that $\langle \chi^2 \rangle \gg \kappa_p^2$
and $\xi_\chi \ll 1$,
we find that the four-point function reduces to
\begin{eqnarray}
\langle \zeta({\vec x_1}) \zeta({\vec x_2}) \zeta({\vec x_3}) \zeta({\vec x_4}) \rangle
& \approx & \frac{{(I_2-I_0)}^2}{I_1^4} 
\left( \langle \zeta ({\vec x_1})\zeta ({\vec x_2}) \rangle
\langle \zeta ({\vec x_2})\zeta ({\vec x_4}) \rangle
\langle \zeta ({\vec x_3})\zeta ({\vec x_4}) \rangle +11~{\rm perms.} \right) \nonumber \\
&&+ \frac{(I_3-3I_1)}{I_1^3} \left( \langle \zeta ({\vec x_1})\zeta ({\vec x_2}) \rangle
\langle \zeta ({\vec x_1})\zeta ({\vec x_3}) \rangle
\langle \zeta ({\vec x_1})\zeta ({\vec x_4}) \rangle +3~{\rm perms.}
\right), \label{preheating-4p}
\end{eqnarray}
where
\begin{equation}
I_3 = \sum_p A_p \epsilon_p \eta_p^3 e^{{-\eta_p^2 \over 2}}.
\end{equation}
We see that the four-point function is contributed by two distinct parts.
Indeed, this is exactly the same form as the standard local type four-point
function which is parametrically represented as
\begin{eqnarray}
\langle \zeta({\vec x_1}) \zeta({\vec x_2}) \zeta({\vec x_3}) \zeta({\vec x_4}) \rangle
&=& \tau_{\rm NL}
\left( \langle \zeta ({\vec x_1})\zeta ({\vec x_2}) \rangle
\langle \zeta ({\vec x_2})\zeta ({\vec x_4}) \rangle
\langle \zeta ({\vec x_3})\zeta ({\vec x_4}) \rangle +11~{\rm perms.} \right) \nonumber \\
&&+ \frac{54}{25}g_{\rm NL} \left( \langle \zeta ({\vec x_1})\zeta ({\vec x_2}) \rangle
\langle \zeta ({\vec x_1})\zeta ({\vec x_3}) \rangle
\langle \zeta ({\vec x_1})\zeta ({\vec x_4}) \rangle +3~{\rm perms.}
\right),
\end{eqnarray}
where $\tau_{\rm NL}$ and $g_{\rm NL}$ are non-linearity parameters for the 
four-point function \cite{Byrnes:2006vq}.
Comparing this with Eq.~(\ref{preheating-4p}), we can read 
\begin{equation}
\tau_{\rm NL}=\frac{{(I_2-I_0)}^2}{I_1^4}=\frac{36}{25} f_{\rm NL}^2,~~~~~
\frac{54}{25} g_{\rm NL}=\frac{(I_3-3I_1)}{I_1^3}. \label{ex-g}
\end{equation}
Our results show that the non-linearity parameters are given by
the combination of the moments $I_i (i=0,1,2,3)$.
Each moment can be numerically evaluated provided position $\eta_p$,
amplitude $A_p$ and width $\epsilon_p$ of each spike are known,
which we will carry out in a following way.
But before going to the numerical results, we can make a crude numerical
estimate of the non-linearity parameters.
From the figure in \cite{Bond:2009xx}, for the relevant range of $\chi+\chi_0$,
we find that $A_p={\cal O}(10^{-5})-{\cal O}(10^{-4})$ and
$\epsilon_p ={\cal O}(10^{-2})-{\cal O}(0.1)$ and $N_s={\cal O}(10)$ spikes.
As is clear from the expression of $I_0$ (see Eq.~(\ref{def-I0})), 
only the spikes having $|\eta_p| \lesssim 1$ contribute to $I_0$ and we have
$I_0 \sim \langle A_p \rangle \langle \epsilon_p \rangle N_s \sim {\cal O}(10^{-6})$.
Similarly, we expect $I_2 \sim {\cal O}(10^{-6})$.
On the other hand, since terms in the summation of $I_1$ are either positive or negative
with the (almost) equal probability, they may partially cancel out each other.
As a result, we would expect $|I_1|$ is at most ${\cal O}(10^{-6})$.
Therefore, unless accidental cancellation to have $|I_2-I_0|/I_0 \ll 1$ occurs, 
as a crude estimation, we have $|f_{\rm NL}| \gtrsim {\cal O}(10^6)$.
Similar consideration yields $|g_{\rm NL}| \gtrsim {\cal O}(10^{12})$.
As we will see below, this estimation indeed provides the correct orders of magnitude
of the non-linearity parameters. 

\begin{figure}[t]
  \begin{center}{
    \includegraphics[scale=1.5]{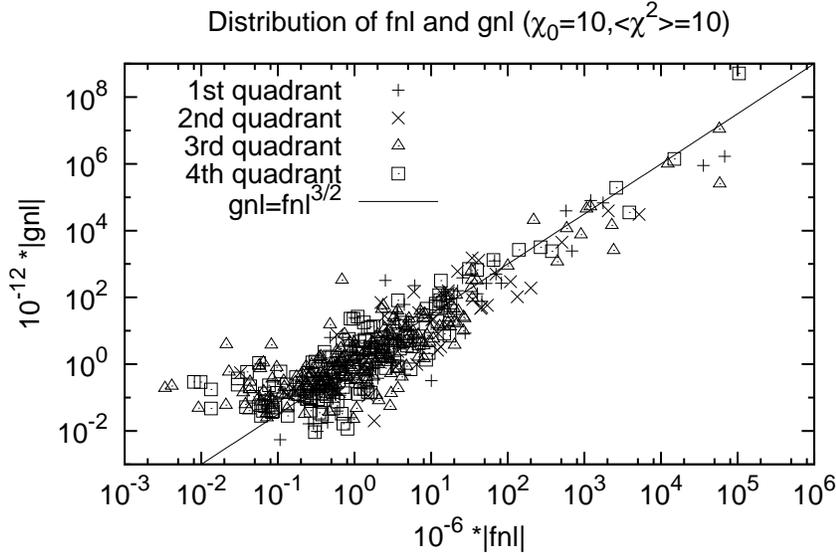}
    }
  \end{center}
  \caption{Distribution of the non-linearity parameters $f_{\rm NL}$ and $g_{\rm NL}$ for
  the case $\chi_0=10$ and $\langle \chi^2 \rangle=10$ in a unit of $10^{-7}M_{\rm Pl}$. 
  The number of realizations is $200$. Since both $f_{\rm NL}$ and $g_{\rm NL}$ can be
  both positive and negative and scatter over a several orders of magnitude,
  both axes are presented logarithmically. Each quadrant is defined for $(f_{\rm NL},g_{\rm NL})$
  plane. For instance, the 2nd quadrant corresponds to a region in which 
  $f_{\rm NL}<0,~g_{\rm NL}>0$.}
 \label{fnl-gnl-distribution}
\end{figure}

\begin{figure}[t]
 \begin{minipage}{0.5\hsize}
  \begin{center}
   \includegraphics[width=80mm]{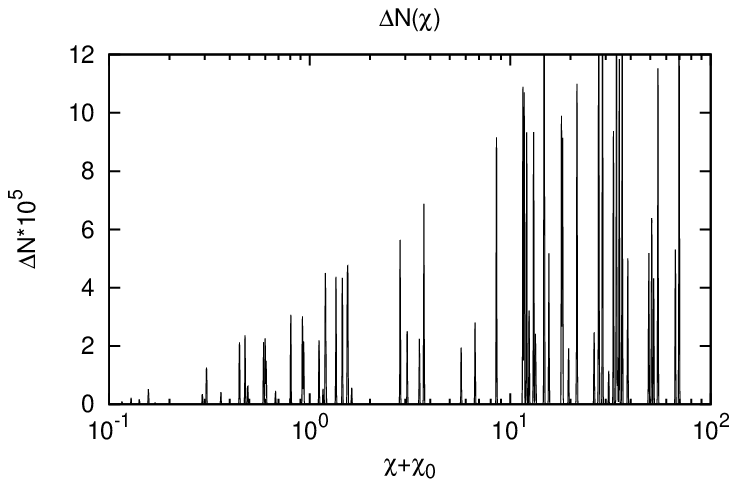}
  \end{center}
  \caption{The function $f(\chi)$ for one realization that yields $10^{-6}\times f_{\rm NL}=0.0055$.}
  \label{fig:small-fnl}
 \end{minipage}
 \begin{minipage}{0.5\hsize}
  \begin{center}
   \includegraphics[width=80mm]{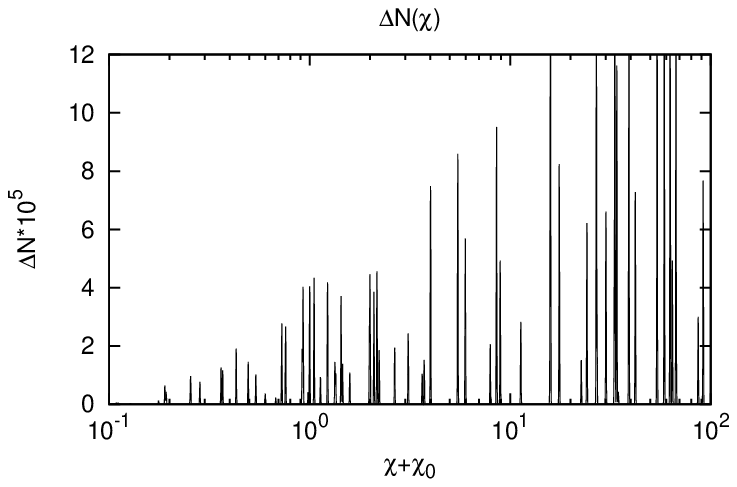}
  \end{center}
  \caption{The function $f(\chi)$ for one realization that yields $10^{-6} \times f_{\rm NL}=1.1 \times 10^5$.}
  \label{fig:large-fnl}
 \end{minipage}
\end{figure}

With the same parameters, we performed $200$ realizations of the mock $f(\chi)$
and calculated $f_{\rm NL}$ and $g_{\rm NL}$ for each realization.
Fig.~\ref{fnl-gnl-distribution} shows the resultant distribution of the
non-linearity parameters plotted in logarithmic scale for fixed parameters 
$\chi_0=10,~\langle \chi^2 \rangle=10$.
Since both $f_{\rm NL}$ and $g_{\rm NL}$ can be positive or negative,
the distributions are shown for each quadrant defined for $(f_{\rm NL},g_{\rm NL})$ plane.
We see that the distribution is highly concentrated in the region
specified by 
$10^{-6}\times f_{\rm NL}={\cal O}(0.1)-{\cal O}(10)$ and
$10^{-12}\times g_{\rm NL}={\cal O}(0.1)-{\cal O}(100)$,
which roughly agrees with our earlier analytic estimation.
Also, we see a clear tendency that larger $|f_{\rm NL}|$ entails larger
$|g_{\rm NL}|$.
We also find that the distribution is wide and covers a several orders of magnitude.
The large $|f_{\rm NL}|$ is realized when $I_1$ becomes much tinier compared to
$I_0$ and $I_2$ due to the accidental cancellation among terms in the summation 
given by Eq.~(\ref{def-I0}).
Since both $f_{\rm NL}$ and $g_{\rm NL}$ have $I_1$ in their denominators,
$g_{\rm NL}$ generically becomes large as well when large $|f_{\rm NL}|$ is realized.
Actually, the distribution for higher $|f_{\rm NL}|$ lies on a line 
$|g_{\rm NL}|=|f_{\rm NL}|^{3/2}$, 
which supports the explanation due to the cancellation of $I_1$. 
The opposite case in which $|I_2-I_0|/I_0 \ll 1$ accidentally happens is also possible
and $f_{\rm NL}$ becomes small in such a case.
Since the numerator of $g_{\rm NL}$, which is given by $I_3-3I_1$, 
is independent of that of $f_{\rm NL}$,
the vanishing of $|I_2-I_0|$ does not result in the vanishing of $I_3-3I_1$ in general,
which explains the disappearance of the positive correlation that exists 
for higher $|f_{\rm NL}|$.
This consideration suggests that, in principle,
either of any very large $|f_{\rm NL}|$ (and $|g_{\rm NL}|$)
or very small $f_{\rm NL}$ can be realized after
we perform sufficiently large number of realizations.
In other words, we can say that the non-linearity parameters are 
quite sensitive to the realizations of $f(\chi)$.
This would imply that the (slightly) imperfect modelling of $f(\chi)$ 
has a possibility of leading to a significant error on the estimation
of the resulting non-linearity parameters. 
Just for illustration, in Figs.~\ref{fig:small-fnl} and \ref{fig:large-fnl},
we show two different $f(\chi)$ due to the different realizations with the same
parameters $\chi_0=10,~\langle \chi^2 \rangle =10$. 
These correspond to the extreme cases where $10^{-6}f_{\rm NL}=0.0055$ 
for the former case and $10^{-6}f_{\rm NL}=1.1 \times 10^5$ for the latter.
It is easy to recognise that they have the similar feature but different in detail,
but not easy, without implementing the numerical calculations, 
to say if they produce quite different $f_{\rm NL}$ or not.

Very large $f_{\rm NL}$ corresponding to right-hand region of Fig.~\ref{fnl-gnl-distribution}
should be taken with caveat.
As we mentioned earlier, such very huge $f_{\rm NL}$ is realized when $I_1$ becomes tiny.
In such a case, the higher order terms in $\xi_\chi$ appearing in the correlation functions, 
which are neglected as being subdominant,
are no longer suppressed.
In order to clarify this point more quantitatively, let us extend the expansion of Eqs.~(\ref{2-point-5})
and (\ref{3-point-2}) to one more higher order;
\begin{eqnarray}
\langle \zeta ({\vec x}) \zeta ({\vec y}) \rangle 
&=&I_1^2 \xi_\chi(r)+\frac{1}{2}{(I_2-I_0)}^2\xi_\chi^2(r)+{\cal O}(\xi_\chi^3), \label{2-point-higher} \\
\langle \zeta({\vec x}_1) \zeta({\vec x}_2) \zeta({\vec x}_3)\rangle
&=&(I_2-I_0)I_1^2 \left( \xi_\chi(r_{12}) \xi_\chi(r_{13})+ 2~{\rm perms.} \right) \nonumber \\
&&+\frac{1}{2} I_1 (I_3-3I_1) (I_2-I_0) (\xi_\chi (r_{12})\xi_\chi^2 (r_{23})+5~{\rm perms.}) \nonumber \\
&&+{(I_2-I_0)}^3 \xi_\chi (r_{12})\xi_\chi (r_{23})\xi_\chi (r_{13}) + {\cal O}(\xi_\chi^3). 
\end{eqnarray}
We find from the first equation that the higher order terms become important
if we have $I_1^2 \lesssim \frac{1}{2} {(I_2-I_0)}^2 \xi_\chi \simeq 10^{-12} \xi_\chi$.
In terms of $f_{\rm NL}$, we can rewrite this condition as $f_{\rm NL} \gtrsim 10^6/\xi_\chi$.
Therefore, if $f_{\rm NL}$ becomes larger than $10^6/\xi_\chi$, the standard expression
of the three-point function written as the two product of the two-point functions, 
which is given by Eq.~(\ref{3-fnl}), 
loses its validity and the use of the constant $f_{\rm NL}$ to characterize the
strength of non-Gaussianity becomes pointless.
In stead of the standard relation, in the limit $I_1 \to 0$, we have 
\begin{equation}
{\langle \zeta({\vec x}_1) \zeta({\vec x}_2) \zeta({\vec x}_3)\rangle}^2=
8 \langle \zeta({\vec x}_1) \zeta({\vec x}_2) \rangle 
\langle \zeta({\vec x}_2) \zeta({\vec x}_3) \rangle
\langle \zeta({\vec x}_3) \zeta({\vec x}_1) \rangle.
\end{equation}
Thus the three-point function is completely determined by the two-point function
without unknown parameter.

\begin{figure}[t]
  \begin{center}{
    \includegraphics[scale=1.5]{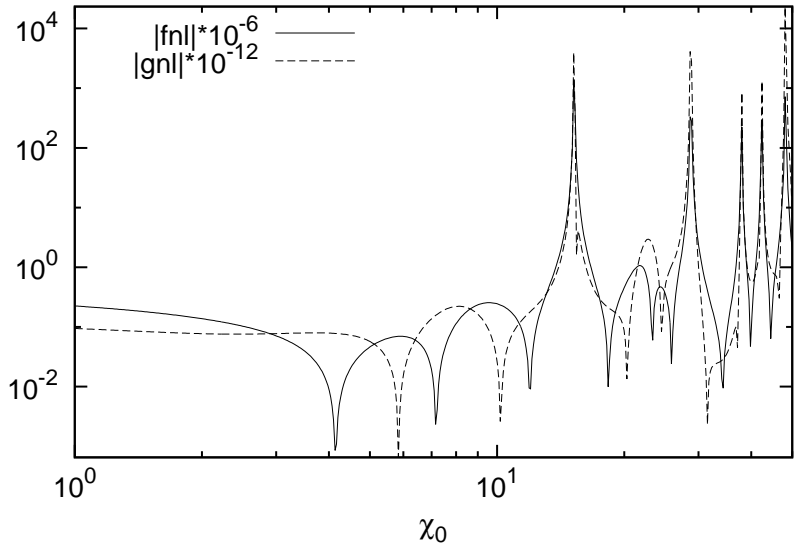}
    }
  \end{center}
  \caption{Dependence of the non-linearity parameters on $\chi_0$ for
  one realization with $\langle \chi^2 \rangle =10$
  in a unit of $10^{-7}M_{\rm Pl}$. The vertical axis represents $10^{-6}~|f_{\rm NL}|$
  and $10^{-12}~|g_{\rm NL}|$ in logarithmic scale.
  }
 \label{chi0-fnl-gnl}
\end{figure}

We also study how the non-linearity parameters vary as we shift $\chi_0$
for a fixed $f(\chi)$ generated randomly (one realization). 
Fig.~\ref{chi0-fnl-gnl} shows the result in which $\chi_0$ is varied
from $1$ to $50$ with the condition $\langle \chi^2 \rangle =10$.
As we can see, $f_{\rm NL}$ becomes quite tiny (and even becomes zero) 
at some specific values of $\chi_0$. 
$g_{\rm NL}$ also does, but at different values of $\chi_0$,
which is expected because of the reasoning we made earlier.
We also observe that $f_{\rm NL}$ is significantly enhanced (and even diverge)
at some specific values of $\chi_0$ for which $g_{\rm NL}$ is also amplified as well.
This can be again understood as the accidental vanishing of $I_1$
due to the cancellation of terms entering the summation of $I_1$.
Although this is for one realization, we checked that the qualitative feature
remains the same for other realizations.
Thus, it is a general consequence that we have $f_{\rm NL}={\cal O}(10^6)$ and
$g_{\rm NL}={\cal O}(10^{12})$ for most of the values of $\chi_0$, but they happen 
to become either tiny or huge if $\chi_0$ is suitably fine-tuned.

To summarize our findings, we have two sources of the uncertainties
that deaden the predictability of the model on the non-linearity parameters.
The first one is due to the incomplete imitation of the mock $f(\chi)$.
As we have seen, two mock $f(\chi)$ that look similar occasionally yield quite different
values of $f_{\rm NL}$ and $g_{\rm NL}$ with some small probability.
This issue may be, in principle, resolved by knowing the correct form of the real $f(\chi)$ precisely,
which is beyond the scope of this paper.
Furthermore, even if we know the correct form of $f(\chi)$, 
it is not clear of whether the Gaussian fitting given by Eq.~(\ref{fitting-f}),
which is essential in the sense that it allows analytic study of the correlation functions
to some extent, is approximate enough to derive the reliable values of the non-linearity parameters.
This is a difficult problem and we do not pursue it in this paper.
Instead, we take a conservative stance that our approach can only estimate the likely values 
of $f_{\rm NL}$ and $g_{\rm NL}$.

The second uncertainty is due to the indeterminacy of $\chi_0$,
which is more fundamental than the first one.      
Since $\chi_0$ is a statistical quantity reflecting the large wavelength modes
outside the box size, it is impossible in principle
to predict a definite value of $\chi_0$.
As a result, we cannot predict the values of $f_{\rm NL}$ and $g_{\rm NL}$, 
although it may be allowed to say at least that $f_{\rm NL}={\cal O}(10^6)$ and
$g_{\rm NL}={\cal O}(10^{12})$ are likely to happen.

\subsection{Consistency check of the mock mapping}
\label{comparison}
\begin{figure}[t]
 \begin{minipage}{0.5\hsize}
  \begin{center}
   \includegraphics[width=80mm]{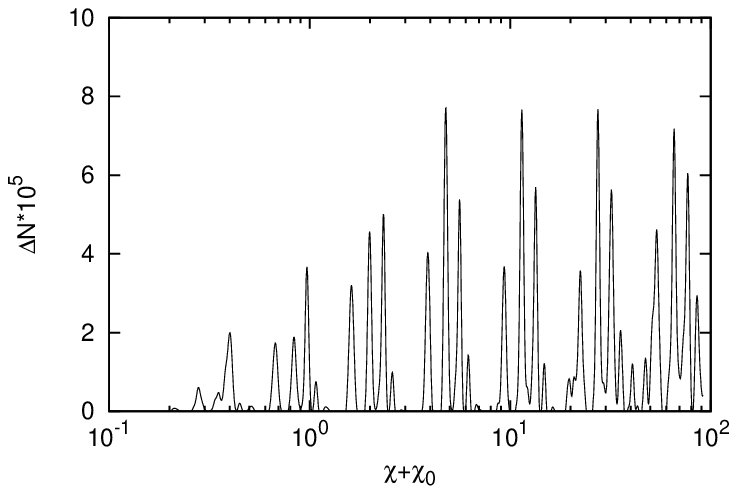}
  \end{center}
  \caption{Smoothed $f(\chi)$ of Fig.~1 presented in \cite{Bond:2009xx}.
  Unit of the horizontal axis is $10^{-7}M_{\rm Pl}$.}
  \label{simulation-deltaN-chi}
 \end{minipage}
 \begin{minipage}{0.5\hsize}
  \begin{center}
   \includegraphics[width=80mm]{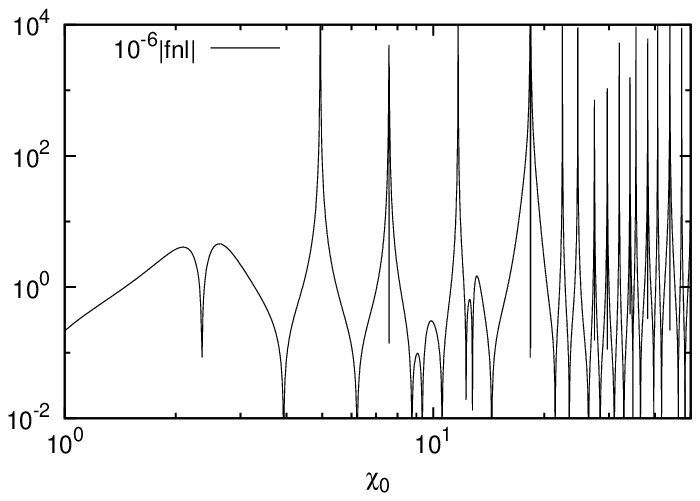}
  \end{center}
  \caption{Dependence of $f_{\rm NL}$ on $\chi_0$ with $\langle \chi^2 \rangle =10$
  in a unit of $10^{-7}M_{\rm Pl}$. The vertical axis represents $10^{-6}~|f_{\rm NL}|$
  in logarithmic scale.}
  \label{simulation-chi0-fnl}
 \end{minipage}
\end{figure}

So far, our study is done with the mock mappings which are not the ones
obtained by the lattice calculations.
It is therefore interesting to compare the results based on the mock mappings with
the ones based on the lattice calculations.
Since we do not have the original numerical data of $f(\chi)$, 
we have read off the coordinates of $f(\chi)$ out of the Fig.~1 presented in \cite{Bond:2009xx}
by using the public software.
In order to extract the position, height and width of each spike, 
we smoothed the obtained $f(\chi)$ by the Gaussian window function with a width $0.015 (\chi_0+\chi)$
which is smaller than $\sqrt{\langle \chi^2 \rangle}$ but is large enough to eliminate the fine spikes.
Fig.~\ref{simulation-deltaN-chi} shows the resultant smoothed graph.
Height of each spike is a little bit smaller than the original one given in \cite{Bond:2009xx},
which is caused by the smoothing.
The width is, instead, more broadened to compensate the loss of the height.  
From this data, we extract the position, height and width of each spike.
We set the minimum height of spike we count in the analysis to be $10^5 \Delta N$.
We choose this value to avoid the inclusion of small spikes which appear to deviate
from normal distribution function.

Fixing $\chi_0$ and $\langle \chi^2 \rangle$, we can calculate the moments $I_0,~I_1,\cdots$
by using the Eqs.~(\ref{def-I0}) and so on.
Fig.~\ref{simulation-chi0-fnl} shows the dependence of $f_{\rm NL}$ on $\chi_0$ with 
$\langle \chi^2 \rangle =10$ by using the formula (\ref{ex-fnl}).
We see that Fig.~\ref{simulation-chi0-fnl} looks quite similar to Fig.~\ref{chi0-fnl-gnl}
in a sense that $f_{\rm NL}$ is typically ${\cal O}(10^6)$ but accidentally becomes very 
small or large for some particular values of $\chi_0$.
This supports that our findings derived in the previous section by using the mock mappings
capture the essential part and are not modified significantly by the use of the real mapping.

\subsection{Observational consequences}
So far, our analysis is based on the assumption that the total curvature
perturbation originates from preheating.
This cannot describe the primordial perturbation of our Universe since the
produced curvature perturbation from preheating does not have enough amplitude
to explain the observed amplitude.
Assuming $I_1=10^{-6}$ and $\xi_\chi =0.04$ corresponding to $r/L=10^{-2}$ as typical values, 
we have $\langle \zeta ({\vec x}) \zeta ({\vec y}) \rangle =4\times 10^{-14}$
from Eq.~(\ref{2-point-5}).
At the Sachs-Wolfe region corresponding to low $\ell$ in the CMB multipole $\ell \lesssim 60$,
the curvature perturbation is related to the CMB temperature anisotropy by
$\zeta ({\vec x})=-5 \Delta T ({\vec x})/T_{\rm CMB}$,
where $T_{\rm CMB} \approx 2.7~{\rm K}$ is the CMB mean temperature.
Using the WMAP observations which detect $\approx 10^3~{\rm \mu K}^2$ of
the power of the temperature fluctuations \cite{Bennett:2012fp}, we find
$\langle \zeta ({\vec x}) \zeta ({\vec y}) \rangle \approx 7 \times 10^{-10}$
is required from observations.
This is about four orders of magnitude larger than that from preheating,
which is observed in Fig.~\ref{two-point-function}.
Therefore, we need additional perturbations that account for the dominant
part of the total curvature perturbations.
Let us assume that the standard adiabatic perturbation from inflation
plays this role.
With this assumption, the total curvature perturbation is a mixture
of two contributions, which can be written as
\begin{equation}
\zeta=\zeta_{\rm inf}+\zeta_{\rm pre}.
\end{equation}
It should be now understood that all the calculations done in the previous 
subsections are for $\zeta_{\rm pre}$.
It is known that the single field inflation with the canonical kinetic term 
predicts deviation of $\zeta_{\rm inf}$ from Gaussianity is suppressed by
the slow-roll parameters.
If the Lagrangian of the inflaton field is more complex or there are multiple
fields responsible for inflation,
non-Gaussianity of $\zeta_{\rm inf}$ can be boosted up to observable level in general
(see for instance \cite{Komatsu:2009kd} and references therein.). 
In this section, we assume that $\zeta_{\rm inf}$ is Gaussian just for simplicity.
From the requirement that $\zeta$ explains the observed amplitude of
the CMB temperature anisotropy, we have
\begin{equation}
\langle \zeta ({\vec x}) \zeta ({\vec y}) \rangle \approx 
\langle \zeta_{\rm inf} ({\vec x}) \zeta_{\rm inf} ({\vec y}) \rangle.
\end{equation}
Therefore, to fit this with the observed slope of the two-point function (spectral index)
is achieved simply by picking up the suitable inflaton potential.
Dynamics of preheating does not enter this game.
On the other hand,
from the Gaussianity ansatz for $\zeta_{\rm inf}$, we also have
\begin{eqnarray}
\langle \zeta ({\vec x_1}) \zeta ({\vec x_2}) \zeta ({\vec x_3}) \rangle \approx 
\langle \zeta_{\rm pre} ({\vec x_1}) \zeta_{\rm pre} ({\vec x_2}) \zeta_{\rm pre} ({\vec x_3}) \rangle.
\label{mix-3-point}
\end{eqnarray}
In the region where the lowest order approximation given by Eq.~(\ref{3-point-2})
works well, Eq.~(\ref{mix-3-point}) can be written as
\begin{equation}
\langle \zeta ({\vec x_1}) \zeta ({\vec x_2}) \zeta ({\vec x_3}) \rangle
\approx s^2 \frac{I_2-I_0}{I_1^2} \left( 
\langle \zeta ({\vec x_1}) \zeta ({\vec x_2}) \rangle \langle \zeta ({\vec x_1}) \zeta ({\vec x_3}) \rangle
+2~{\rm perms.} \right), \label{mix-3-point-2}
\end{equation}
where $s$ is defined by
\begin{equation}
s \equiv \frac{\langle \zeta_{\rm pre}({\vec x})\zeta_{\rm pre}({\vec y}) \rangle}{\langle \zeta({\vec x})\zeta({\vec y}) \rangle}, \label{def-s}
\end{equation}
and represents the relative contribution of the curvature perturbation
of the preheating origin to the total two-point function.
By definition, we have $s \le 1$.
Strictly speaking, $s$ is a function of $|{\vec x}-{\vec y}|$ (although its
dependence on $r$ is logarithmic) and the factorization of Eq.~(\ref{def-s}) is 
mathematically inconsistent.
Nonetheless, this factorization is still useful to estimate the magnitude of the three-point
function if it is observationally important or not. 
Taking $\langle \zeta_{\rm pre} \zeta_{\rm pre}\rangle \approx 4\times 10^{-14}$
and $\langle \zeta \zeta \rangle \approx 7\times 10^{-10}$ as representative values,
a typical value of $s$ is estimated as $s =6\times 10^{-5}$.
Comparison of Eq.~(\ref{mix-3-point-2}) with Eq.~(\ref{3-fnl}) enables us to
define the effective $f_{\rm NL}$ given by
\begin{equation}
\frac{6}{5}f_{\rm NL}=s^2 \frac{I_2-I_0}{I_1^2}.
\end{equation}
Choosing $s$ to be $6\times 10^{-5}$, we see that the effective $f_{\rm NL}$ gets smaller
than the original one, for instance, shown in Figs.~\ref{fnl-gnl-distribution}
and \ref{chi0-fnl-gnl}, by a small number $s^2=4\times 10^{-9}$.
Interestingly enough, while the original $f_{\rm NL}$ is too large to
be compatible with the existing observational constraint $|f_{\rm NL}| \lesssim 100$
except for the cases in which $\chi_0$ is fine-tuned to some specific values,
the suppression by the factor makes the effective $f_{\rm NL}$ to be typically
in the range ${\cal O}(10^{-3})-{\cal O}(0.1)$ \footnote{
The general relativistic second order effects produce $f_{\rm NL}={\cal O}(1)$.
Such effects are not considered in this paper.
},
which is below the detectable amplitude by observations.
One may wonder if one can boost $f_{\rm NL}$ to much higher value than ${\cal O}(0.1)$ 
by making $I_1$ be accidentally small.
However, this is not so since $f_{\rm NL}$ in the mixed case is proportional to
$I_1^2$ rather than to $I_1^{-2}$.
Thus, smaller $I_1$ results in smaller $f_{\rm NL}$ and 
$f_{\rm NL}\lesssim {\cal O}(1)$ is the robust upper bound on $f_{\rm NL}$ in
the massless preheating of our case.

In a similar way, the four-point function can be written as
\begin{eqnarray}
\langle \zeta ({\vec x_1}) \zeta ({\vec x_2}) \zeta ({\vec x_3}) \zeta ({\vec x_4}) \rangle & \approx & 
\langle \zeta_{\rm pre} ({\vec x_1}) \zeta_{\rm pre} ({\vec x_2}) 
\zeta_{\rm pre} ({\vec x_3}) \zeta_{\rm pre} ({\vec x_4}) \rangle \nonumber \\
&=& \frac{{(I_2-I_0)}^2}{I_1^4} s^3
\left( \langle \zeta ({\vec x_1})\zeta ({\vec x_2}) \rangle
\langle \zeta ({\vec x_2})\zeta ({\vec x_4}) \rangle
\langle \zeta ({\vec x_3})\zeta ({\vec x_4}) \rangle +11~{\rm perms.} \right) \nonumber \\
&&+ \frac{(I_3-3I_1)}{I_1^3} s^3\left( \langle \zeta ({\vec x_1})\zeta ({\vec x_2}) \rangle
\langle \zeta ({\vec x_1})\zeta ({\vec x_3}) \rangle
\langle \zeta ({\vec x_1})\zeta ({\vec x_4}) \rangle +3~{\rm perms.}
\right). \label{mix-4-point}
\end{eqnarray}
Correspondingly, the non-linearity parameters are given by
\begin{equation}
\tau_{\rm NL}=\frac{{(I_2-I_0)}^2}{I_1^4}s^3=\frac{36}{25s} f_{\rm NL}^2,~~~~~
\frac{54}{25} g_{\rm NL}=\frac{(I_3-3I_1)}{I_1^3}s^3. \label{mix-tnl-gnl}
\end{equation}
Thus, $\tau_{\rm NL}$ is enhanced by $s^{-1}={\cal O}(10^4)$ compared with $f_{\rm NL}^2$. 
Actually, $\tau_{\rm NL}$ is always larger than or equal to $\frac{36}{25}f_{\rm NL}^2$ irrespective of 
the underlying model \cite{Suyama:2007bg,Smith:2011if,Sugiyama:2012tr}.
Particularly, as the above equation shows, enhancement of $\tau_{\rm NL}$ 
compared to $f_{\rm NL}^2$ is the general feature for the case where
the curvature perturbation is a mixture of the dominating Gaussian perturbation
and the sub-dominant non-Gaussian perturbation \cite{Ichikawa:2008iq,Ichikawa:2008ne}.
In light of the situation that Planck satellite can constrain $\tau_{\rm NL}$ 
to be up to $\sim 600$ \cite{Kogo:2006kh}, 
observational studies of trispectrum can be quite useful to examine if 
the trace of preheating is left in the non-Gaussianity of the curvature perturbation.
The second equation of (\ref{mix-tnl-gnl}) shows that we may expect $g_{\rm NL}$
to be ${\cal O}(1)$.
This is a several orders of magnitude smaller than the sensitivity expected
to be accomplished in future observations \cite{Smidt:2010ra}, 
but is still larger by a few orders of magnitude than the one coming 
from the standard adiabatic fluctuation $\zeta_{\rm inf}$. 

\section{General cases}
\label{Gc}

Here, we have seen that, for the curvature perturbations
sourced by preheating, three-point and four-point functions
at the leading order in $\xi_\chi$ take exactly the same forms as those 
of the standard local type non-Gaussianity for which the curvature perturbation
is given by the expansion of $\chi$ as
\begin{equation}
\zeta ({\vec x})=\sum_{n=0} \frac{1}{n!} N_n \chi^n({\vec x})-{\rm average}, \label{g-local}
\end{equation}
at the leading order in $\xi_\chi$.
In this section, we will show that this is not the special property limited 
to the curvature perturbations from preheating, 
but the generic feature of any curvature perturbation given by Eq.~(\ref{def-zeta}),
for which expansion in terms of $\chi$ needs not to be a good approximation, 
provided $\chi$ obeys Gaussian statistics.
Thus, the only assumption we will make in this section is that $\xi_\chi$ is small 
enough to allow us to pick up the leading terms in any correlator of $\zeta$.
We will thus leave the function form of $f(\chi)$ unspecified.

In the present situation, we can Taylor-expand Eq.~(\ref{N-point}) in
terms of $\xi_\chi$ truncate it at the leading order.
Thus the resultant expression consists of terms, 
each of which contains the product of $\xi_\chi (r_{ij})$.
Each fully connected term in the correlator, 
which we are interested in,
corresponds to the product such that any ${\vec x_i} ~(i=1,\cdots,N)$
appears at least once in the argument of $\xi_\chi$ and a corresponding diagram,
which has $N$-vertices to each of which ${\vec x_i}$ is attached without duplication
and is constructed by drawing line between ${\vec x_i}$ and ${\vec x_j}$
if we have $r_{ij}$ in $\xi_\chi$,
is simply connected.
For instance, $\xi_\chi (r_{12}) \xi_\chi^2 (r_{34})$ appearing in the 
four-point function does not give the connected term, while
$\xi_\chi (r_{12}) \xi_\chi (r_{23}) \xi_\chi (r_{34})$ does.
At the leading order in $\xi_\chi$, corresponding connected diagrams
are tree diagrams, which do not have any loop inside, because
any connected loop diagram is made by attaching lines to some connected 
tree diagram with the same number of vertices, which is accompanied by
additional powers of $\xi_\chi$.
It turns out that any connected tree diagram having $N$-vertices has
$N-1$ internal lines.
This means that any leading connected term in the $N$-point function of $\zeta$
contains a product of $(N-1) ~\xi_\chi$ 's.
Therefore, at the leading order in $\xi_\chi$, we have 
\begin{equation}
\langle \zeta ({\vec x_1}) \cdots \zeta({\vec x_N}) \rangle =
{\left( -\langle \chi^2 \rangle \right)}^{N-1} 
{\hat {\cal C}}\bigg[ \prod_{p=1}^{N-1} \xi_\chi (r_{p_i,p_j})
\int \left( \prod_{i=1}^N \frac{d\sigma_i}{2\pi}~f_{\sigma_i} 
e^{-\frac{\langle \chi^2 \rangle}{2} \sigma_i^2} \right) 
\sigma_{p_i} \sigma_{p_j} \bigg], \label{general-1}
\end{equation}
where ${\hat {\cal C}} [\cdots ]$ means we take only connected parts of $[\cdots ]$. 
As mentioned above, we can associate each term in Eq.~(\ref{general-1}) with the 
corresponding connected diagram.
Since all the possible combinations $(\cdots,p_i,p_j,\cdots)$ in the product of $\xi_\chi$ 's
appear in Eq.~(\ref{general-1}),
all the possible connected tree diagrams contribute to Eq.~(\ref{general-1}).
Since the connected tree diagrams having $N$-vertices that are not isomorphic to 
each other show different scale dependence due to different appearance of the combination
of the product of $\xi_\chi(r_{p_i,p_j})$.
The number of connected tree diagrams is equal to that of the independent parameters required to parametrize
the $N$-point function.
This is exactly the same property for the $N$-point function of the standard local 
type curvature perturbation of the form (\ref{g-local}) at the tree level \cite{Byrnes:2007tm,Yokoyama:2008by}.  
Therefore, Eq.~(\ref{general-1}) has exactly the same structure as the one for
the standard local type curvature perturbation.

Conversely, from a given connected tree diagram, we can construct a corresponding 
non-linearity parameter contributing to Eq.~(\ref{general-1}) according to a following rule.
For a vertex to which a vector ${\vec x_i}$ is attached, we assign 
$-\frac{\langle \chi^2 \rangle}{2\pi}f_{\sigma_i} e^{-\frac{\langle \chi^2 \rangle}{2} \sigma_i^2} \sigma_i^{g_i}$,
where $g_i$ is a number of lines coming out of the vertex.
Without a loss of generality, we can assume $g_1 \le g_2 \cdots \le g_N$.
For a line connecting the vertices ${\vec x_i}$ and ${\vec x_j}$, 
we assign $\xi_\chi (r_{ij})$. 
We then multiply all of them and integrate it over $(\sigma_1,\cdots,\sigma_N)$,
which gives one term in Eq.~(\ref{general-1}).
In order to make things simpler,
let us define the moments $J_n~(n=0,1,\cdots)$ by
\begin{equation}
J_n = -\langle \chi^2 \rangle \int \frac{d\sigma}{2\pi}~
f_\sigma e^{-\frac{\langle \chi^2 \rangle}{2} \sigma^2} \sigma^n. \label{def-Jn}
\end{equation}
A connected tree diagram having $N$-vertices can be characterized by
a sequence consisting of $N$ integers $(g_1,\cdots,g_N)$.
Therefore, a non-linearity parameter $f(g_1,\cdots,g_N)$ associated with the corresponding diagram
can be written as
\begin{equation}
f(g_1,\cdots,g_N)=J_1^{-2(N-1)}\prod_{i=1}^N J_{g_i}.
\end{equation}
As illustrations, let us consider the three-point function.
There is only one connected tree diagram with three vertices
characterized by $(g_1,g_2,g_3)=(1,1,2)$.
Then, the non-linearity parameter can be written as
\begin{equation}
\frac{6}{5}f_{\rm NL}=J_1^{-2} J_2.
\end{equation}
It can be explicitly verified that this equation exactly recovers
Eq.~(\ref{ex-fnl}) for the preheating case.
In a similar way, for the four-point function,
there are two connected tree diagrams characterized by $(1,1,2,2)$
and $(1,1,1,3)$.
The former case gives $\tau_{\rm NL}$ given by
\begin{equation}
\tau_{\rm NL}=J_1^{-4} J_2^2=\frac{36}{25}f_{\rm NL}^2.
\end{equation}
The latter case give $g_{\rm NL}$ given by
\begin{equation}
\frac{54}{25} g_{\rm NL}=J_1^{-3} J_3.
\end{equation}
This also coincides with Eq.~(\ref{ex-g}) for the preheating case.

Finally, let us show that just as in the case for massless preheating in 
the last section, $J_n$ is related to the $n$-th derivative of the smoothed
e-folding number.
The Fourier transform of the Gaussian window function defined by Eq.~(\ref{gauss-window})
can be written as
\begin{equation}
{\tilde W_g}(\sigma)=\int d\chi~W_g (\chi) e^{-i\chi \sigma}
=e^{-\frac{\langle \chi^2 \rangle}{2}\sigma^2}.
\end{equation}
With this Fourier-transformed window function, we can write the smoothed
e-folding number $f^R(\chi)$ as
\begin{equation}
f^R(\chi)=\int d\chi' \int \frac{d\sigma}{2\pi} f_\sigma e^{i(\chi+\chi')\sigma} W_g (\chi')=
\int \frac{d\sigma}{2\pi} ~f_\sigma e^{i\chi \sigma} 
e^{-\frac{\langle \chi^2 \rangle}{2}\sigma^2}.
\end{equation}
Thus, its $n$-th derivative becomes
\begin{equation}
\frac{d^n f^R}{d\chi^n}\bigg|_{\chi=0}=
i^n \int \frac{d\sigma}{2\pi} f_\sigma e^{-\frac{\langle \chi^2 \rangle}{2}\sigma^2} \sigma^n
=-\frac{i^n}{\langle \chi^2 \rangle} J_n,
\end{equation}
where we have used Eq.~(\ref{def-Jn}) in the last equation.
This shows that, to compute the correlation function of $\zeta$, 
we can apply the standard diagrammatic rule following from 
the $\delta N$ formalism, namely to assign $n$-th derivative of the
e-folding number to the vertex where $n$ lines are attached, 
simply by replacing the bare e-folding number with the smoothed e-folding number.

\section{Conclusion}
There are some inflationary models where the resultant curvature perturbation $\zeta$
is a function of the Gaussian scalar field but its dependence cannot be approximated
by quadratic expression.
Representative models include the trigonometric mapping and the massless preheating models.
In this paper, we provided a general formulation to calculate the correlation functions
of the curvature perturbation which is an arbitrary function of the Gaussian scalar field $\chi$.
The $N$-point function of the curvature perturbation is written as $N$-dimensional
integral, which can be evaluated once the Fourier transform of the mapping between
$\zeta$ and $\chi$ field is given.
We applied the formalism first to the common local type non-Gaussian curvature perturbation
and verified that it reproduces the standard local form of the three-point function.
We then considered more non-trivial cases including the model in which 
the curvature perturbation is the trigonometric function of the scalar field.
Due to the appearance of the Dirac's $\delta$-function, this model allows analytic
evaluation of the correlation function.

By using the fitting formula provided in \cite{Bond:2009xx} which approximates each spike
of the mapping $\zeta (\chi)$ as the normal distribution function,
we could analytically perform the integrals for the two-point, three-point and four-point functions of $\zeta$.
For physically relevant scales, 
forms of the correlation functions reduce to the standard local type ones.
In particular, while the non-linearity parameters defining the strength of the non-Gaussianity
of $\zeta$ are related to the derivatives of the e-folding number with
respect to $\chi$ in the standard case, the role of the derivatives is replaced
by the moments defined as sums over spikes with suitable weights for the case of preheating,
or equivalently, by the derivatives of the e-folding number smoothed in the field space.
Due to the lack of the knowledge of the precise positions of the spikes and their
widths and amplitudes,
we randomly generated the mapping of mock $\zeta (\chi)$ 's having the similar properties
to the original one found in \cite{Bond:2009xx} and studied how
the resultant non-linearity parameters differ according to different realizations.
We found that the non-linearity parameters take naively estimated values for
most of the realizations but some realizations yield very large or small values
due to occasional cancellation among contributions from each spike. 
Even after the parameters appearing in the original Lagrangian are fixed,
statistical indeterminacy of the average value of $\chi$ in our observable universe
also causes another uncertainty of the non-linearity parameters. 
We therefore studied dependence of the non-linearity parameters on the average
value of $\chi$ for a fixed realization.
It was found that for some specific values (these values differ by different
realizations) the non-linearity parameters are sharply enhanced or diminished
compared to the naively estimated values.
These analyses show that we can only estimate the likely values (order of magnitude)
of the non-linearity parameters.

Due to the insufficient amplitude of the curvature perturbation from preheating,
the curvature perturbation in our observable universe, 
if the massless preheating indeed happened in the early universe,
must be a mixture of another dominant component and the subdominant one of 
preheating origin.
Assuming the dominant component to be the standard adiabatic Gaussian
perturbation coming from inflaton,
the mixture reduces the non-Gaussianity of the total curvature perturbation
to the observationally allowed range.
The trispectrum is relatively amplified compared to the bispectrum and
use of both observables will be useful to search preheating signature buried
in the curvature perturbation.

By a diagrammatic approach, 
we also showed that when the correlation function of $\chi$ can be treated  
perturbatively, the forms of the correlation functions of $\zeta$
coincide with the standard local type ones at the leading order approximation.
While the non-linearity parameters are given by the product of the derivatives of the e-folding number
in the standard case,
they are given exactly the same manner even in our case provided we replace the bare e-folding
number with the one smoothed in the field space with a Gaussian window function.
Diagrammatic rule for converting the diagram to the corresponding non-linearity
parameter was also provided.
\\

\noindent {\bf Acknowledgments:} 
The authors thank to Masahiro Kawasaki and Jun'ichi Yokoyama for helpful discussions.
We also thank to Liyi Gu for suggesting the use of the public software which extracts data from graphs.
TS thanks the Leung Center for Cosmology and Particle Astrophysics (LeCosPA), 
National Taiwan University for the kind hospitality during his visit when this paper is completed.
This work was supported by Grant-in-Aid for JSPS Fellows No.~1008477 (TS)
and No. 24-2775(SY).

\bibliographystyle{unsrt}
\bibliography{draft}

\end{document}